\documentclass[pra,twocolumn,aps,10pt]{revtex4-1}

\usepackage{amsmath,amssymb,amsthm}

\usepackage{graphicx,xcolor}

\usepackage[colorlinks=true]{hyperref}

\usepackage[T1]{fontenc}
\usepackage{xparse}
\usepackage{bbm}

\newcommand{\ii}{\mathrm{i}}
\let \Re \relax
\DeclareMathOperator{\Re}{Re}
\let \Im \relax
\DeclareMathOperator{\Im}{Im}
\DeclareMathOperator{\tr}{tr}
\DeclareMathOperator{\arcosh}{arcosh}
\DeclareMathOperator{\sgn}{sgn}
\DeclareMathOperator{\sinc}{sinc}

\newcommand{\barC}{\bar{C}}

\NewDocumentCommand{\chainst}{}{\textsf{(St)}}
\NewDocumentCommand{\chainsp}{}{\textsf{(Sp)}}
\NewDocumentCommand{\chainfl}{}{\textsf{(Fl)}}

\newcommand{\mk}{\mkern1mu}

\begin{document}

\title{Topological origin of quantized transport in non-Hermitian Floquet chains}

\author{Bastian H{\"o}ckendorf}
\author{Andreas Alvermann}
\email{alvermann@physik.uni-greifswald.de}
\thanks{Corresponding author}
\author{Holger Fehske}
\affiliation{Institut f{\"u}r Physik, Universit{\"a}t Greifswald, Felix-Hausdorff-Str.~6, 17489 Greifswald, Germany}

\begin{abstract}
We show that non-Hermiticity enables
topological phases with unidirectional transport in one-dimensional Floquet chains.
The topological signatures of these phases are non-contractible loops in the spectrum of the Floquet propagator that are separated by an imaginary gap. Such loops occur exclusively in non-Hermitian Floquet systems.
We define the corresponding topological invariant as the winding number of the Floquet propagator relative to the imaginary gap.
To relate topology to transport,
we introduce the concept of
regularized dynamics of non-Hermitian chains.
We establish that, under the conditions of regularized dynamics, transport is quantized in so far as the charge transferred over one period equals the topological winding number.
We illustrate these theoretical findings 
 with the example of a Floquet chain that features a topological phase transition and
 acts as a charge pump in the non-trivial topological phase. 
 We finally discuss whether these findings justify the notion that non-Hermitian Floquet chains support topological transport.
\end{abstract}

\maketitle

\section{Introduction}

Quantum Hall systems~\cite{Klitzing, TKNN} and topological insulators~\cite{KaneMelePRL, Konig2007} are manifestations of a fundamental connection between topology and transport.
Topological transport is characterized by two properties:
It is quantized and robust~\cite{HasanKane2010, RevModPhys.83.1057}.
Ultimately, quantization and robustness
are consequences of the bulk-boundary correspondence, which relates transport via chiral (or helical) boundary states to the topological properties of an insulating bulk~\cite{TKNN,KaneMelePRL}.
Importantly, topological transport requires boundary or surface states.
One-dimensional systems can exhibit non-trivial topology~\cite{PhysRevLett.42.1698, Kitaev_2001},
but do not support robust transport without additional assumptions~\cite{PhysRevB.27.6083,RevModPhys.82.1959,PhysRevLett.120.106601}.

Recent research has shown that non-Hermiticity considerably extends this picture~\cite{PhysRevLett.120.146402, PhysRevLett.106.213901, Hararieaar4003, PhysRevLett.123.206404, PhysRevLett.124.056802,PhysRevB.98.245130,PhysRevA.98.042120}.
While the new non-Hermitian topological phases, with imaginary and point gaps in addition to the real gaps of the Hermitian case~\cite{PhysRevX.8.031079, PhysRevB.99.235112, PhysRevX.9.041015}, have been classified for static systems, conclusive results on non-Hermitian topological transport are still rare.
Even the status of a non-Hermitian bulk-boundary correspondence remains debatable,
since boundary transport can be modified 
outside of the constraints 
the correspondence imposes on
 Hermitian systems~\cite{PhysRevLett.121.086803, PhysRevLett.121.136802, PhysRevLett.123.190403, PhysRevB.99.201103}.
Notably, non-Hermiticity is not a theoretical construct but appears naturally in,
e.g., acoustics~\cite{PhysRevLett.121.124501}, electronics~\cite{2019arXiv190711562H}, or optics and photonics~\cite{SzameitJPB,RevModPhys.91.015006, Regensburger2012, Weimann2016, PhysRevLett.115.040402,bandres2018}.

The subject of this work is the relation between topology and transport in one-dimensional
non-Hermitian chains. We show that, contrary to the Hermitian case,
these chains can act as charge pumps.
If---but only if---we consider Floquet chains with a time-periodic Hamiltonian $H(t+T) = H(t)$, we can even observe transport that is quantized and robust.
Whether quantization and robustness allow us to call this ``topological transport''  is a question that will be reconsidered in the conclusions.

The paper is organized as follows:
In Sec.~\ref{sec:topology} we explain the different topological scenarios realized in non-Hermitian chains.
We identify a topological phase that occurs exclusively in non-Hermitian Floquet systems, and lies outside of the established classification~\cite{PhysRevX.8.031079, PhysRevB.99.235112, PhysRevX.9.041015} for static non-Hermitian systems.
Sec.~\ref{sec:transport} studies transport in non-Hermitian chains,
which we relate to topological properties via the concept of regularized dynamics introduced in Sec.~\ref{sec:RegularizedDynamics}.
Sec.~\ref{sec:RobustTransport} investigates the robustness of transport in non-Hermitian Floquet chains, in particular under the influence of disorder.
Our findings are summarized in the conclusions in Sec.~\ref{sec:conclusion}. 
Additional details and computations that are not essential for the understanding of the main concepts are given in the Appendices.

\begin{figure}
\includegraphics[width=\columnwidth]{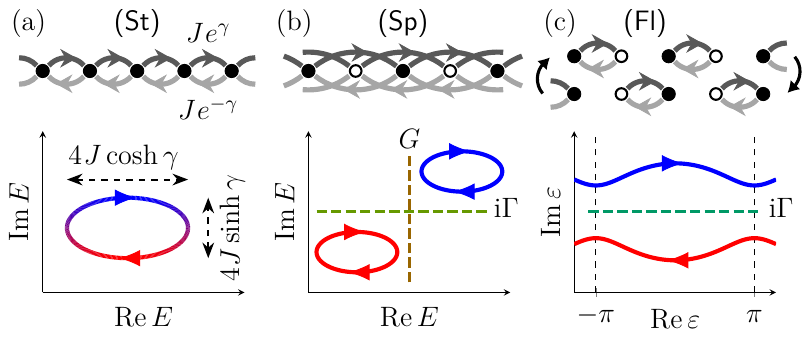}
\caption{Conceptual overview of the topological scenarios for non-Hermitian chains.
Panel (a): The dispersion of a static chain \textsf{(St)} with directional nearest-neighbor hopping
is an elliptical loop in the complex energy plane.
Panel (b): By splitting the hopping spatially,
 the chain \textsf{(Sp)} supports two loops separated by a real ($G$) and imaginary ($\ii \Gamma$) line gap.
Panel (c):
By splitting the hopping temporally,
the Floquet chain \textsf{(Fl)} supports loops that traverse the complex quasienergy zone
with its $\varepsilon \mapsto \varepsilon + 2\pi$ periodicity.
}
\label{fig:overview}
\end{figure}

\section{Topology of non-Hermitian chains}
\label{sec:topology}

An overview of the different topological scenarios for non-Hermitian chains 
is given in Fig.~\ref{fig:overview}, using simple generic models as illustrative examples.
In all examples, the parameter $J$ specifies the strength, the parameter $\gamma$ the directionality of hopping. 

\subsection{Static chains}

We start with the standard example of a non-Hermitian chain \textsf{(St)} with directional hopping~\cite{PhysRevX.8.031079, 2019arXiv190711562H} (see also App.~\ref{app:ChainSt}).
As a function of momentum $k$, the spectrum $E(k) = J (e^{-\ii k + \gamma} +  e^{\ii k - \gamma})$ of the 
Hamiltonian
\begin{equation}
H_\mathsf{St} = J \sum_{n \in \mathbb Z} e^\gamma |n +1\rangle\langle n| \!+  e^{-\gamma} |n\rangle\langle n+1|
\end{equation}
is an ellipse
with real (imaginary) semi-axis $2 J \cosh \gamma$ ($2 J \sinh \gamma$).
The elliptical loop $k \mapsto E(k)$ is contractible, and thus topologically trivial.
Non-trivial topology can be enforced by defining that a point gap exists inside of the loop~\cite{PhysRevX.8.031079, PhysRevB.99.235112, PhysRevX.9.041015}, but this definition of non-trivial topology is not related to notions of transport.

For the chain \textsf{(Sp)} we split the hopping spatially,
which doubles the unit cell such that the spectrum contains two elliptical loops
(see App.~\ref{app:ChainSp} for the full definition of the Hamiltonian).
These loops can be separated with a staggered potential $\pm \Delta \in \mathbb C$, leading to a 
real ($G$) or imaginary ($\ii \Gamma$) line gap, but remain contractible. 

\subsection{Floquet chains}
The situation changes qualitatively if we split the hopping temporally to obtain the Floquet chain \textsf{(Fl)}.
One period (of length $T \equiv 1$) in this chain comprises two alternating directional hopping steps, with Hamiltonian
\begin{equation}
H^{(1)}_\mathsf{Fl} = J \sum_{n \in \mathbb Z} e^\gamma |2 n+1\rangle\langle 2n| +  e^{-\gamma} |2n\rangle\langle 2n+1|
\end{equation}
in the first and 
\begin{equation}
H^{(2)}_\mathsf{Fl} = J  \sum_{n \in \mathbb Z} e^\gamma |2 n\rangle\langle 2n-1| +  e^{-\gamma} |2n-1\rangle\langle 2n|
\end{equation}
in the second half-period (see App.~\ref{app:ChainFl} for an extended specification).
For a Floquet chain, we must consider quasienergies 
$\varepsilon(k)$ which, in contrast to the energies
$E(k)$ of a static chain, are determined only up to multiples of $2 \pi$.
The multi-valuedness allows for the two loops $k \mapsto \varepsilon_{1,2}(k)$ of the Floquet chain \textsf{(Fl)} in Fig.~\ref{fig:overview},
which wrap around the quasienergy zone and thus are non-contractible.
Note that the loops appear with opposite chirality $\varepsilon_{1,2}(k+2 \pi) =\varepsilon_{1,2}(k) \pm 2 \pi$.
We will now identify these non-contractible loops
as the signatures 
of the topological phase of a non-Hermitian Floquet chain,
and later also as the origin of quantized transport.

\subsection{Non-contractible loops in the Floquet spectrum}

\begin{figure}
\includegraphics[width=\columnwidth]{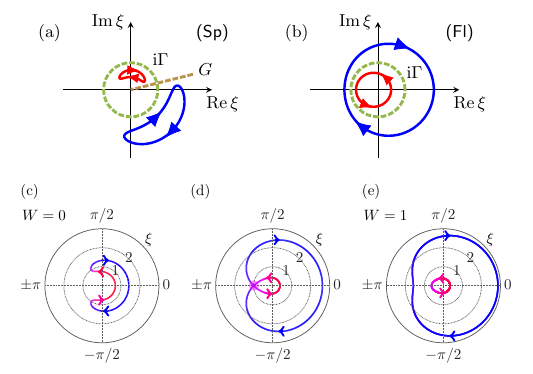}
\caption{
Panels (a), (b):
Spectrum of the (Floquet) propagator 
 for the chains \textsf{(Sp)},~\textsf{(Fl)} from Fig.~\ref{fig:overview}.
Only the Floquet chain \textsf{(Fl)} can exhibit non-contractible loops with non-zero winding number.
Panels (c)--(e): Topological phase transition in the Floquet chain \textsf{(Fl)} with $J=\pi/3$.
The transition at  $\gamma = \gamma_c \approx 0.55$ (panel (d)) separates the trivial ($\gamma = 0.2$, panel (c))
from the non-trivial ($\gamma =0.6$, panel (e)) phase.
}
\label{fig:spectrum}
\end{figure}

To illustrate the specific topology of  the one-dimensional non-Hermitian setting we consider the spectrum of the Floquet-Bloch propagator $\hat U(k) \equiv U(T,k)$ (see Fig.~\ref{fig:spectrum}),
which is the solution of the Schr\"odinger equation $\ii \partial_t U(t,k)=H(t,k) U(t,k)$ after one period
$t=T$.
The eigenvalues $\xi_m(k)=e^{-\ii \varepsilon_m(k)}$ of $\hat U(k)$ lie in the punctured complex plane $\mathbb C \setminus \{0\}$.
Exclusion of the origin results from invertibility $U(t)^{-1} = U(-t)$ of the propagator, which also holds in the non-Hermitian setting.

Note that static chains can be embedded into the Floquet picture by choosing an artificial period $T$. 
Since the Floquet propagator of a static chain is
$U = \exp(-\ii \mkern1mu T H)$,
quasienergies and energies are in one-to-one correspondence $\varepsilon_m(k) \equiv T E_m(k) \mod 2 \pi$ for sufficiently small $T$, as long as $\Re E_m(k) \in (-\pi/T, \pi/T)$.

Fig.~\ref{fig:spectrum} provides us with three topological insights.
First, non-contractible loops of a Floquet chain wind around the origin.
Conceptually, the origin serves as a natural point gap for the Floquet propagator (but \emph{not} for the Hamiltonian).
Second, an imaginary
gap $\ii \Gamma$ partitions the spectrum into an inner and outer part,
separated by the circle $\phi \mapsto e^{-\ii \phi + \Gamma}$.
A non-trivial imaginary gap
requires a non-Hermitian Hamiltonian,  where the spectrum of the propagator is not restricted to the unit circle by unitarity.
Third, a real gap $G$, which corresponds to a radial line $r \mapsto r e^{-\ii G}$ ($r \in \mathbb R_+$),
prohibits non-contractible loops and implies trivial topology.

We here observe
a fundamental difference between 
static and Floquet chains.
For static chains (without symmetries),
real and imaginary gaps are equivalent via multiplication $H \mapsto \ii H$ of the Hamiltonian by the imaginary unit~$\ii$.
For Floquet chains, real and imaginary gaps are strictly inequivalent.
In consequence, non-Hermitian Floquet chains can support topological phases that do not appear in static chains.

\subsection{Topological invariant: The winding number}

Translating the topological concepts of Fig.~\ref{fig:spectrum} into an invariant,
we are led to 
the $\mathbb Z$-valued winding number 
\begin{equation} \label{W1}
\begin{aligned}
W(\Gamma)
&=\frac{\ii}{2\pi} \sum_{e^\Gamma < |\xi_m|} \,  \int_{-\pi}^{\pi}   \xi_m(k)^{-1} \,\partial_k \mkern1mu \xi_m(k) \, \mathrm{d}k \\
&= \frac{1}{2\pi}\sum_{\Gamma < \Im \varepsilon_m} \big[ \mathrm{Re} \,  \varepsilon_m(k) \big]_{k=-\pi}^{k=\pi}  \;.
\end{aligned}
\end{equation}
$W(\Gamma)$ is defined with respect to an imaginary gap $\ii \Gamma$.
Only eigenvalues $\xi_m(k)$ outside (or quasienergies $\varepsilon_m(k)$ above) the gap contribute.
A non-contractible loop in clockwise direction contributes with a positive integer.
Note that we normalize the Brillouin zone to $k \in (-\pi, \pi]$, independently of the size of the unit cell.

The trivial imaginary gap $\Gamma = - \infty$,
where the entire spectrum of $U$ contributes, gives the total winding number $W(-\infty) = 0$.
This result follows because the spectrum of $U$ cannot move through the origin,
such that non-contractible loops appear with opposite chirality.

Hermitian chains, where loops cannot be separated by imaginary gaps,
as well as static non-Hermitian chains, which have only contractible loops,
necessarily have zero winding number $W(\Gamma)=0$ for all $\Gamma$.

In contrast, non-zero winding numbers are possible in non-Hermitian Floquet chains. The 
chain \textsf{(Fl)}
features a topological phase transition 
at the critical value
$\gamma_c = \arcosh \, (1/\sin |J|)$
(see App.~\ref{app:ChainFl} for a derivation from the eigenvalues of the Floquet propagator).
As shown in Fig.~\ref{fig:spectrum}, the spectrum below the transition ($|\gamma| < \gamma_c$) consists of a single loop with periodicity $k \mapsto k + 4 \pi$. The winding number  is zero.
At the transition ($|\gamma| = \gamma_c$), 
the spectrum possesses an exceptional point at $k=0$.
Starting from the exceptional point, the spectrum splits into two loops
above the transition ($|\gamma| > \gamma_c$).
The loops occur with opposite chirality, and are separated by an imaginary gap $\ii \Gamma$ at $\Gamma = 0$.
The associated winding number is non-zero, with $W(\Gamma) = 1$ for $\gamma > 0$ (as in Fig.~\ref{fig:spectrum}) and $W(\Gamma) =-1$ for $\gamma < 0$.

Although the appearance of
non-contractible loops in the spectrum of the propagator
is reminiscent of
the anomalous phase of two-dimensional Hermitian Floquet insulators~\cite{KitagawaPRB, Rudner, Nathan, HockendorfJPA, HockendorfPRB, HAF19, Maczewsky, Mukherjee, Peng2016, GreifRostock}, the topological phase observed here
is specific to  one-dimensional non-Hermitian Floquet chains.
Formally, it requires a non-trivial imaginary gap. Physically, the non-contractible loops that appear here are not associated with boundary states as in the Floquet insulator, but with the spectrum of the infinite chain.

\section{Transport in non-Hermitian chains}
\label{sec:transport}

Transport in non-Hermitian chains can be quantified with the charge
\begin{equation}\label{charge}
C(n) = \mathrm{tr}_\mathbb{Z} \, (U^\dagger [P_n,U])
\end{equation}
transferred over one period through a fictitious layer between sites $n-1$ and $n$ from the left to the right of the chain (see App.~\ref{app:TransferredCharge} for an extended derivation).
Here, $P_n$ is the projection onto sites $i \ge n$, that is
$P_n |i \rangle = |i\rangle$ for $i \ge n$ and $P_n |i\rangle = 0$ for $i < n$. 
Eq.~\eqref{charge} generalizes standard expressions (e.g.,~\cite{Graf2018}) for $C(n)$ to the non-Hermitian setting.

\subsection{Transport and the Floquet spectrum}

For a Hermitian chain, evaluation of the trace shows that the total charge transfer is $C(n) = 0$.
This is only true in dimension one: The analogous expression in two dimensions gives the charge transferred by chiral boundary states, which, of course, can be non-zero~\cite{Graf2018, PhysRevB.97.195312}.

\begin{figure}
\includegraphics[width=\columnwidth]{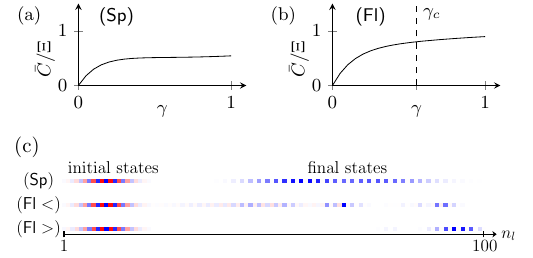}
\caption{
Panels (a), (b): Transferred charge $\bar C$ in the chains \textsf{(Sp)}, \textsf{(Fl)}, as a function of $\gamma$ (with $J=\pi/3, \Delta = 3 \ii$).
The curve is normalized with the factor $\Xi = \max_\xi |\xi|^2$.
Hermiticity (at $\gamma=0$) enforces $\bar C = 0$.
Panel (c): Probing transport with wave packet propagation in the chains \textsf{(Sp)}, \textsf{(Fl)}
(with $\gamma = 0.8, 0.09, 1.5$ from top to bottom, and $n_p=40$ periods).
}
\label{fig:charge}
\end{figure}

For a non-Hermitian chain, whether static or Floquet, $C(n)\!\ne 0$ becomes possible (see Fig.~\ref{fig:charge}).
However, allowing for non-Hermiticity involves intrinsic complications for the physical interpretation of transport.
First, current is no longer conserved but a charge $c(n) = \langle n| [U,U^\dagger] |n \rangle$ can accumulate
at site $n$.
To get rid of the resulting site dependence of $C(n)$, we average the transferred charge $\bar C = (1/L) \sum_{n=0}^{L-1} C(n)$ over a unit cell of $L$ sites of a translationally invariant chain.

We can now derive the momentum space expression
$\bar C = \frac{\ii}{2\pi} \int_{-\pi}^{\pi}  \tr_\mathbb{L} \big(\hat U^\dagger(k) \, \partial_k \hat U(k)\big) \mathrm{d}k$,
where the trace $\tr_\mathbb{L}$ runs over the unit cell (see App.~\ref{app:MomentumSpace}).
The second complication in comparison to the Hermitian setting is that this expression depends explicitly on the eigenvectors of $\hat U(k)$, and is not invariant under unitary transformations.
This is not an artefact of the derivation, but a fundamental consequence of the fact that non-Hermiticity allows for non-orthogonal eigenvectors of $\hat U(k)$.

If, however, the eigenvectors of $\hat U(k)$ are orthogonal (a requirement equivalent to the normality condition $[\hat U(k), \hat U(k)^\dagger]=0$ known from linear algebra~\cite{GolubVanLoan}),
\begin{equation}\label{IbarNormal}
\bar  C \; \underset{\text{(orth)}}{=} \;
        \frac{\ii}{2\pi} \sum_{m=1}^L  \int_{-\pi}^{\pi}   \xi_m(k)^* \, \partial_k \xi_m(k) \, \mathrm{d}k
\end{equation}
is given entirely in terms of the eigenvalues $\xi_m(k)$ of $\hat U(k)$.
Note that it has a geometric interpretation as the area (with orientation) enclosed by the eigenvalue loops $k \mapsto \xi_m(k)$.

\subsection{Transferred charge and wave-packet propagation}

The transferred charge can be measured by real-space propagation of wave packets, which would be the method of choice, e.g., in photonic waveguide systems.
In wave packet propagation, the transferred charge gives the average propagation distance, weighted by the norm of the propagated wave packets, which can deviate from unity in the non-Hermitian setting.

Fig.~\ref{fig:charge} shows that, without further provisions, the transferred charge $\bar C$ 
does not reveal much about the nature of transport in the chains \textsf{(Sp)},~\textsf{(Fl)} .
However, probing transport by means of wave packet propagation 
we observe a significant difference: 
While wave packets usually spread out during propagation, 
they propagate almost without spreading in the Floquet chain \textsf{(Fl)} above the topological phase transition (row ``\textsf{(Fl$\,>$)}'' in Fig.~\ref{fig:charge}), where $\bar C \to 1$.

\section{Topology and transport: Regularized dynamics}
\label{sec:RegularizedDynamics}

We should not expect that a strict relation between transport and topology in non-Hermitian chains holds in general situations.
To understand the conditions that are required for a strict relation we introduce the concept of
 regularized dynamics (RD) of non-Hermitian chains. For RD we demand that (i)
the dominant eigenvalues of the Floquet propagator have modulus one, (ii) the modulus of all other eigenvalues is infinitesimally close to zero, and (iii) the eigenvectors are mutually orthogonal.
Then, Eq.~\eqref{IbarNormal} can be used to compute $\bar C$.
Since $\xi^* = \xi^{-1}$ for $|\xi|=1$, this equation reduces to Eq.~\eqref{W1} for the winding number, if $W(\Gamma)$ is computed for an imaginary gap $-\infty < \Gamma < 0$,
that separates the eigenvalues with modulus zero ($\Im \varepsilon \to -\infty$) from those with modulus one ($\Im \varepsilon=0$). We thus obtain the fundamental relation
\begin{equation}\label{Fund}
 \bar C \; \underset{\text{(RD)}}{=} \; W(\Gamma)
\end{equation}
between transport and topology in non-Hermitian chains with RD.
In particular, the transferred charge $\bar C$ 
is quantized, and vanishes in a static 
chain where $W(\Gamma)=0$.

\subsection{Regularized dynamics as a physical limit}

It remains to assess whether RD of non-Hermitian chains is an artificial construction or a relevant concept.
First, it should not be surprising that some assumptions are required for a quantitative relation between transport and topology~\cite{PhysRevLett.120.106601}---the same being true also for Hermitian Floquet insulators where a condition $\hat U(k) \equiv 1$ is imposed~\cite{Rudner}.
The RD of non-Hermitian Floquet chains differs from previous regularization concepts in so far as
(i) RD applies to the Floquet propagator, not to a static Hamiltonian~\cite{PhysRevX.9.041015},
(ii) RD requires an imaginary gap $\ii \Gamma$ instead of a real gap of a unitary Floquet propagator~\cite{Rudner}.

\begin{figure}
\includegraphics[width=\columnwidth]{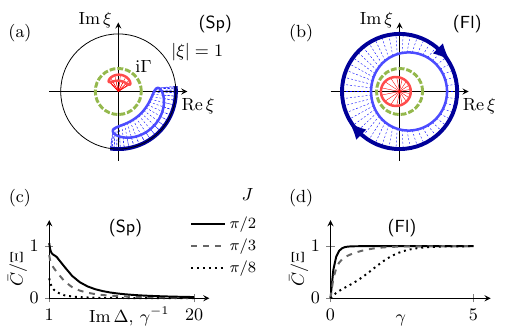}
\caption{
Panels (a), (b): In the RD limit the spectrum of the propagator outside (inside) of the imaginary gap $\ii \Gamma$ moves to the unit circle (origin).
Panels (c), (d):
Transferred charge $\bar C$, for various hopping $J = \pi/2, \pi/3, \pi/8$,
as one approaches the RD limit in the static chain \textsf{(Sp)} (for $\Im \Delta, \gamma^{-1} \to \infty$) or
the Floquet chain \textsf{(Fl)} (for $\gamma \to \infty$).
In this limit, $\bar C = W(\Gamma)$.
}
\label{fig:regularize}
\end{figure}

For a non-Hermitian chain, RD can be achieved either (i) through regularization of the propagator,
or (ii) as a physical limit in parameter regimes with strong damping. 
Regularization of the propagator corresponds to a continuous deformation of $\hat U(k)$
such that the eigenvalues $\xi_m(k)$  move in- or outwards to the origin or the unit circle, as illustrated in Fig.~\ref{fig:regularize}.
As a technical complication of the non-Hermitian setting, also the eigenvectors of the propagator have to be deformed.
Regarding topological properties, details of the regularization procedure are not relevant, as long as the imaginary gap $\ii \Gamma$ stays open during the deformation such that the winding number does not change (two techniques are described in App.~\ref{app:Regularization}).

As a physical limit, RD is realized in parameter regimes where strong damping suppresses some eigenvalues of the Floquet propagator, while the dominant states incur uniform loss or gain.
We can drop a factor $\Xi^{1/2} = \max_\xi |\xi|$, in order to normalize $\hat U(k)$ with a uniform imaginary shift of $H$, such that the dominant states have zero loss.
For the static chain \textsf{(Sp)}, RD is achieved for example in the limit $\gamma \to 0, \Im \Delta \to \infty$ (shown in Fig.~\ref{fig:regularize}). With the interpretation of Eq.~\eqref{IbarNormal} as an area it is evident that $\bar C \to 0$ in the RD limit, in accordance with the fundamental relation~\eqref{Fund}. Note that absence of transport is compatible with a point gap, which can enforce non-contractible eigenvalue loops but does not prevent minimization of the enclosed area by continuous deformation.

For the Floquet chain \textsf{(Fl)}, RD is realized in the limit $\gamma \to 0$, or $\gamma \to \pm \infty$ (shown in Fig.~\ref{fig:regularize}).
For $\gamma \to 0$, we end up below the topological phase transition, 
with a trivial gap $\Gamma=-\infty$. This results in a unitary propagator 
 without directed transport ($\bar C = 0 = W(\Gamma)$).
 For $\gamma \to \pm \infty$, we end up above the topological phase transition.
One non-contractible eigenvalue loop $k \mapsto e^{\mp \ii k}$ survives on the unit circle.
This eigenvalue loop encloses the circle area $\pm 2\pi$,
such that Eq.~\eqref{IbarNormal} gives $\bar C = \pm 1 =W(\Gamma)$ in accordance with the fundamental relation~\eqref{Fund}.

\section{Robustness of transport}
\label{sec:RobustTransport}

From Eq.~\eqref{Fund} the quantization of transport in the RD limit follows.
With regard to our initial characterization of topological transport we should ask for the robustness of the transferred charge $\bar C$ seen in Fig.~\ref{fig:regularize}.

Robustness has two aspects.
 First, we note that $\bar C$ does not depend on the remaining free model parameters (in Fig.~\ref{fig:regularize}, the hopping $J$). Importantly, RD does not require fine-tuning of the system to a point in parameter space, but is realized on a parameter manifold where transport and topology are invariant.

\begin{figure}
\includegraphics[width=\columnwidth]{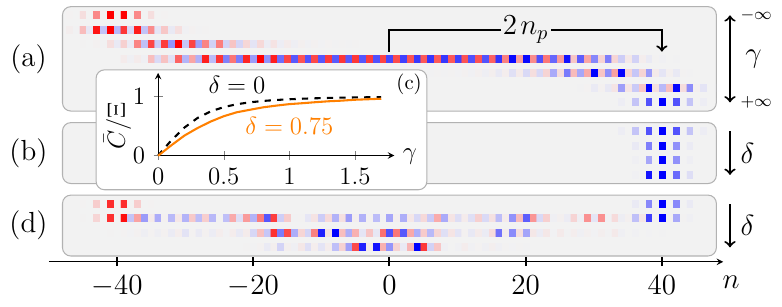}
\caption{The Floquet chain \textsf{(Fl)} as a charge pump:
Shown is a wave packet, centered initially at site $n=0$, after propagation over $n_p = 20$ periods.
Panel (a): Propagation for $\gamma = -\infty, -1, -0.25, 0, 0.25, 1, \infty$
from top to bottom  (with $J=\pi/3$).
Panel (b): Propagation in the RD limit $\gamma \to \infty$, for disorder $\delta = 0, 0.25, 0.5, 0.75$ from top to bottom (with $J=\pi/3$).
Inset panel (c): Charge $\bar C$ transferred per period. 
Panel (d): Propagation at $J=\pi/2$, $\gamma=0$, with  $\delta$ as in panel (b).
}
\label{fig:pump}
\end{figure}

Second, Fig.~\eqref{fig:pump} probes the robustness of transport in the chain \textsf{(Fl)} via wave packet propagation.
Panel (a) shows propagation in the translationally invariant chain.
In the RD limit $\gamma \to \pm \infty$, half of the wave packet (on either blue or red sites) survives propagation. Over $n_p$ periods, the wave packet moves without spreading by $2 \, \bar C  n_p$ sites, with $\bar C = \pm 1$.
 Disorder in panel (b), with hoppings chosen randomly per bond from the interval $[J(1-\delta), J(1+\delta)]$, does \emph{not} affect transport in the RD limit, which is thus revealed to be truly robust.
 Note that the wave packets depends on disorder even for $\gamma \to \infty$,
 but the averaged charge $\bar C$ does not (see inset panel (c)).
Quite differently, panel (d) shows that fine-tuning to `perfect coupling' ($J=\pi/2$, $\gamma=0$), with distinct propagation in the ordered chain, does not survive addition of disorder.

\section{Conclusions}
\label{sec:conclusion}

In conclusion,
we present a theory of topology and transport in non-Hermitian chains.
While transport can occur in any non-Hermitian chain, 
only Floquet chains allow for non-trivial topology.
Therefore, only non-Hermitian Floquet chains might possibly act as topological charge pumps with quantized and robust transport.

Indeed, as our example of the Floquet chain \chainfl{} shows, non-Hermitian Floquet chains possess quantized transport that is robust against parameter variations and disorder, provided that the chain approaches the limit of regularized dynamics.
In view of the fundamental relation~\eqref{Fund} the emergence of quantized transport in the limit of regularized dynamics is not an arbitrary or random observation, but a direct consequence of the existence of non-trivial topological phases in Floquet chains. 
The topological origin of quantized transport is the appearance of non-contractible loops in the Floquet spectrum.

Evidently, non-Hermitian Floquet systems possess topological phases with interesting transport and dynamical properties that are not realized in static or Hermitian systems.
It remains a matter of debate whether the quantized transport that appears (only) in the limit of regularized dynamics deserves the designation of ``topological transport''.
Therefore, we believe that future research on non-Hermitian systems should focus increasingly on the relation between topology and transport.  

Certainly, further investigation of 
situations with disorder, defects, or interfaces is required especially 
with a view towards experiments on non-Hermitian topological systems using, e.g.,
photonic waveguides or electric circuits to implement Floquet instead of static non-Hermitian chains~\cite{2019arXiv190711562H}.
The relevance of such investigations is strongly supported by the recent work of Ref.~\cite{fedorova2019topological} where, fully independently of the present work, the theoretical concept of non-adiabatic charge pumping in non-Hermitian Floquet chains is developed and accompanied with the observation of quantized non-Hermitian transport in plasmonic waveguide arrays.

\appendix

\section{The three chains \chainst{}, \chainsp{}, \chainfl{}}

To specify the Hamiltonians of the three chains \chainst{}, \chainsp{}, \chainfl{} we use the usual bra-ket notation, where $|n\rangle$ denotes the state at site $n \in \mathbb Z$.
For the chains \textsf{(Sp)}, \textsf{(Fl)}, we identify `filled' sites $\bullet$  in Fig.~\ref{fig:overview} with even $n$, and `open' sites $\circ$ with odd $n$.

\subsection{The static chain \chainst{}}\label{app:ChainSt}

The Hamiltonian of the chain $\textsf{(St)}$,
\begin{equation}
 H_\mathsf{St} = J \, \sum_{n \in \mathbb Z} \big( e^{\gamma} |n+1\rangle \langle n| + e^{-\gamma} |n\rangle \langle n+1| \big) \;,
\end{equation}
includes directional hopping between nearest neighbors.
For $\gamma > 0$ ($\gamma <0$) hopping to the right (left) is enhanced.
The Hamiltonian is Hermitian only for $\gamma=0$, when it reduces to that of a tight-binding chain with non-directional hopping.

The Hamiltonian $H_\mathsf{St}$ is invariant under translations by $L=1$ sites, and has a scalar Bloch Hamiltonian
\begin{equation}
\hat H_{\mathsf{St}}(k) = J(e^{-\ii k +\gamma}+e^{\ii k-\gamma}) \;.
\end{equation}
This gives the elliptical energy loop $E_\mathsf{St}(k) \equiv \hat H_{\mathsf{St}}(k)$ shown in Fig.~\ref{fig:overview}.
Note that we can also write
$E_\mathsf{St}(k) = 2 J (\cosh \gamma \cos k - \ii \sinh \gamma \sin k)$.

\subsubsection{\chainst{}: Transferred charge}

The Hamiltonian $H_\mathsf{St}$, and thus also the propagator $U_\mathsf{St}(t)= \exp ( - \ii t H_\mathsf{St} )$,
satisfies the prerequisite $[U^\dagger, U]=0$ of Eq.~\eqref{app:CBarNormal}.
We thus compute, with $\xi(k) = e^{-\ii t E_\mathsf{St}(k)}$,
\begin{equation}\label{barCSt}
\begin{split}
\barC &=  \frac{\ii}{2\pi} \int_{-\pi}^{\pi}   \xi(k)^* \, \partial_k \xi(k) \, \mathrm{d}k \\
&= \frac{t}{2\pi} \int_{-\pi}^{\pi}    \partial_k E_\mathsf{St}(k)  e^{2 t \Im  E_\mathsf{St}(k) }     \, \mathrm{d}k \\
&= - \frac{2 J t \cosh \gamma}{2\pi} \int_{-\pi}^{\pi}   \sin k \, e^{4 J t \sinh \gamma \sin k }     \, \mathrm{d}k
\end{split}
\end{equation}
Using the integral representation of the Bessel functions~\cite{AS70} we get
\begin{equation}\label{barCSt2}
\barC =  2 J t \, \cosh \gamma \, I_1(4 J t \sinh \gamma) \;,
\end{equation}
with the modified Bessel function $I_1(\cdot)$.
We have $\barC=0$ in the Hermitian case $\gamma=0$ (as it must), but the non-Hermitian case $\gamma \ne 0$ allows for $\barC \ne 0$.

\subsubsection{\chainst{}: Regularized dynamics}

The spectrum $E_\mathsf{St}(k)$ of the Hamiltonian $\hat H_{\mathsf{St}}(k)$ consists of a single loop. In the RD limit, the loop should have constant imaginary part, corresponding to constant magnitude $|\xi| = e^{t \Im E}$ of the eigenvalues of the corresponding propagator $U(t)$.
The only way to achieve this is to set $\gamma=0$, recovering a Hermitian chain (possibly with a uniform complex shift of the Hamiltonian).
Therefore, $\bar C = 0$ in the RD limit.

\subsection{The static chain \chainsp{}}
\label{app:ChainSp}

The chain $\textsf{(Sp)}$ essentially consists of two identical copies of the chain $\textsf{(St)}$, placed either on the `filled'  or `open' sites.
To separate the two copies, we include a staggered potential $\Delta \in \mathbb C$. To couple the two copies, we can allow for hopping $\lambda\in \mathbb R$ between the `filled' and `open' sites (although not particularly relevant for the present investigation).
  This results in the Hamiltonian
\begin{equation}\label{app:spham}
\begin{aligned}
 H_\mathsf{Sp} =J\, &\sum_{n \in \mathbb Z} \big( e^{\gamma} |n+2\rangle \langle n| + e^{-\gamma} |n\rangle \langle n+2| \big)
 \\
 +\Delta &\sum_{n \in \mathbb Z} \big (|2n\rangle \langle 2n|-|2n+1\rangle \langle 2n+1|\big)
 \\
 +\lambda &\sum_{n \in \mathbb Z} \big(|n\rangle \langle n+1|+|n+1\rangle \langle n|\big) \; ,
 \end{aligned}
 \end{equation}
 with translational invariance by $L=2$ sites.
 
The corresponding $2 \times 2$ Bloch Hamiltonian is 
\begin{equation}\label{app:spbloch}
\hat H_{\mathsf{Sp}}(k)=\begin{pmatrix} E_{\mathsf{St}}(k)+\Delta & 2\lambda \cos(k/2) \\ 2\lambda \cos(k/2) & E_{\mathsf{St}}(k)- \Delta\end{pmatrix} \;,
\end{equation}
with eigenvalues
\begin{equation}\label{app:sperg}
E_{\mathsf{Sp}}(k)=E_{\mathsf{St}}(k) \pm \sqrt{4\lambda^2\cos^2(k/2) + \Delta^2} \; .
\end{equation}
For sufficiently large $\Delta$ this gives two separated energy loops (for $\Re \Delta = \lambda = 0$, the condition is $|\Im \Delta| > 2 |J \sinh \gamma|$), as sketched in Fig.~\ref{fig:overview}.
We use  $\Delta=3\ii$ and $\lambda = 0.5$ in Fig.~\ref{fig:charge}.

Note that we normalize the Brillouin zone to the interval $k\in(-\pi,\pi]$, independently of the size of the unit cell (hence the factor $k/2$ in Eqs.~\eqref{app:spbloch},~\eqref{app:sperg}).
The eigenvalues $E_{\mathsf{Sp}}(k)$ are $2 \pi$-periodic, while $\hat H_{\mathsf{Sp}}(k)$ acquires an irrelevant phase (cf. the Fourier transform in Eq.~\eqref{eq:Fourier1}).

\subsubsection{\chainsp{}:Transferred charge}

For $\lambda = 0$ the chain \chainsp{} consists of two copies of the chain \chainst{},
such that the transferred charge
\begin{equation}
\barC = 4 J t \, \cosh \gamma \, \cosh  (2 t \Im \Delta) \, I_1(4 J t \sinh \gamma) 
\end{equation}
is the weighted sum of twice Eq.~\eqref{barCSt}, with weighting factors $e^{\pm 2 t \Im \Delta}$ derived from the imaginary part of the staggered potential $\Delta$.
For $\lambda \ne 0$ the analytic computation of $\barC$ becomes tedious, and will not be pursued here.
Numerical data are given in Figs.~\ref{fig:charge},~\ref{fig:regularize}.

\subsubsection{\chainsp{}:Regularized dynamics}

Just as for the chain \chainst{}, the RD limit in the chain \chainsp{} requires $\gamma=0$,
such that $\bar C = 0$.
Now, however, we can separate loops with the staggered potential $\Delta$.
A trivial RD limit is obtained for $\Delta = 0$, reducing the chain to a Hermitian chain.
In the non-trivial RD limit $|\Im \Delta| \to \infty$, there exists one loop below and one loop above the imaginary gap at $\Gamma = 0$. This is the situation sketched in Fig.~\ref{fig:regularize}. Still, we have $\bar C = 0$.

\subsection{The Floquet chain \chainfl{}}
\label{app:ChainFl}

The Hamiltonian of the Floquet chain $\textsf{(Fl)}$
\begin{equation}
 H_\mathsf{Fl}(t) = \begin{cases} 
\, \dfrac2T \, H_\mathsf{Fl}^{(1)}  \quad & \text{if }  n_p  \le t/T <  n_p +\frac12 \;, \\[8pt]
\, \dfrac2T \, H_\mathsf{Fl}^{(2)}             & \text{if }  n_p +\frac12 \le t/T < n_p+1 
 \end{cases}
\end{equation}
 consists of the two alternating steps
\begin{subequations}\label{eq:steps}
\begin{equation}\label{eq:step1}
 H_\mathsf{Fl}^{(1)} = J \sum_{n \in \mathbb Z} \big( e^{\gamma} |2n+1\rangle \langle 2n| + e^{-\gamma} |2n\rangle \langle 2n+1| \big) \; ,
\end{equation}
\begin{equation}\label{eq:step2}
 H_\mathsf{Fl}^{(2)} = J \sum_{n \in \mathbb Z} \big( e^{\gamma} |2n\rangle \langle 2n-1| + e^{-\gamma} |2n-1\rangle \langle 2n| \big)
\end{equation}
\end{subequations}
in each period ($n_p \in \mathbb Z$) of length $T$.
As for the chain \textsf{(Sp)}, $H_\mathsf{Fl}(t)$ is invariant under translations by $L=2$ sites.

Note that we use $T=1$ throughout the present manuscript, both to obtain quasienergies $\varepsilon_m$ from eigenvalues $\xi_m$ via the relation $\xi_m = e^{-\ii \varepsilon_m T} \equiv e^{-\ii \varepsilon_m}$, as well as to interpret static chains as Floquet chains with an artifical period $T$. We do however keep the symbol $T$ wherever suitable to remind us of the meaning of the respective variable, e.g., we write $U(T)$ instead of $U(1)$.

The Floquet propagator $U_{\mathsf{Fl}} \equiv U_{\mathsf{Fl}}(T)$ in real space is
\begin{equation}\label{eq:real_space}
\begin{aligned}
U_{\mathsf{Fl}} =& \;  e^{-\ii H_\mathsf{Fl}^{(2)} T/2}e^{-\ii H_\mathsf{Fl}^{(1)} T/2}\\
=&  \phantom{ {}-{} } c^2 \sum_{n \in \mathbb Z} |n\rangle \langle n|
\\
&-s^2 \sum_{n \in \mathbb Z}\big( e^{2\gamma}|2n+2\rangle \langle 2n|+e^{-2\gamma}|2n-1\rangle \langle 2n+1|\big)
\\
&-\ii sc \sum_{n \in \mathbb Z} \big (e^{\gamma}|2n+1\rangle \langle 2n|+e^{-\gamma}|2n\rangle \langle 2n+1| \big)  
\\
&-\ii sc \sum_{n \in \mathbb Z} \big (e^{\gamma}|2n\rangle \langle 2n-1|+e^{-\gamma}|2n-1\rangle \langle 2n| \big)  
\; ,
\end{aligned}
\end{equation}
with the abbrevations $c\equiv \cos J$, $s\equiv \sin J$.
The Floquet-Bloch propagator in momentum space is
\begin{equation}
\begin{aligned}
\hat U_{\mathsf{Fl}}(k)=\begin{pmatrix} c^2-s^2 e^{-\ii k+2\gamma} & -2\ii sc \cos(k/2+\ii \gamma)\\ -2\ii sc \cos(k/2+\ii \gamma) &  c^2-s^2 e^{\ii k-2\gamma}\end{pmatrix} \; ,
\end{aligned}
\label{FloquetBloch}
\end{equation}
with eigenvalues
\begin{equation}
\label{eq:Fl_eigenvalues}
\begin{aligned}
\xi_{1,2}(k) = &   
\, 1-2 s^2\cos^2(k/2+\ii \gamma)
\\ 
& \pm 2 s \cos(k/2+\ii\gamma) \sqrt{s^2 \cos^2(k/2+\ii \gamma)-1}  \; .
\end{aligned}
\end{equation}
From the eigenvalues we obtain the quasienergies $\varepsilon_{1,2}(k)$ via the relation $\xi_{1,2}(k) = e^{-\ii \varepsilon_{1,2}(k)}$. Eigenvalues and quasienergies are shown in Fig.~\ref{Fig:s1}.

\subsubsection{\chainfl{}: Topological phase transition}

\begin{figure}
\includegraphics[width=1\linewidth]{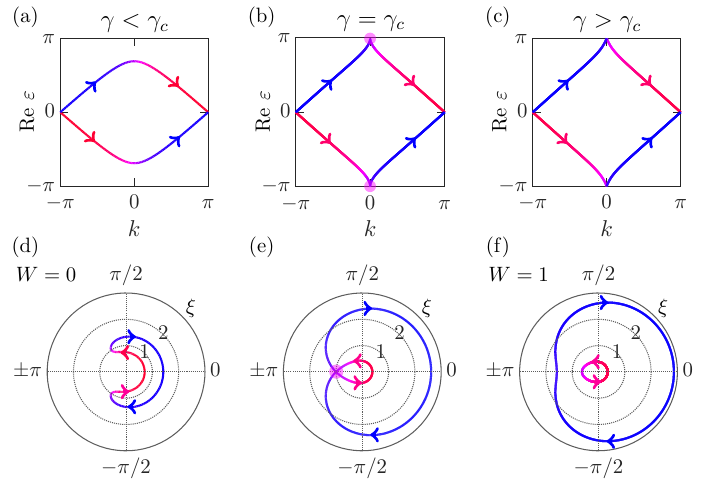}
\caption{Topological phase transition in the Floquet chain \textsf{(Fl)} with $J=\pi/3$, as already shown in Fig.~\ref{fig:spectrum}.
Panels (a)--(c) show the real part $\Re \varepsilon_{1,2}(k)$ of the quasienergies, which have been omitted in Fig.~\ref{fig:spectrum}.
Panels (d)--(f) show the spectrum of eigenvalues $\xi_{1,2}(k)$.
The pink dots in panels (b), (e) indicate the exceptional point at the transition.
}
\label{Fig:s1}
\end{figure}

The topological phase transition  in the chain \chainfl{} occurs when the square root in Eq.~\eqref{eq:Fl_eigenvalues} vanishes,
which happens for $\gamma = \pm \gamma_c$ with the critical value
\begin{equation}
\gamma_c = \mathrm{arcosh}(1/\sin |J|) \;.
\end{equation}
The spectrum of $\hat U_{\mathsf{Fl}}(k)$ consists of a single loop for $|\gamma| < \gamma_c$, 
and of two loops for $|\gamma| > \gamma_c$ (see Fig.~\ref{Fig:s1}). The loops are separated by an imaginary gap at $\Gamma=0$, and have (necessarily) opposite winding number.
Note that at the transition, the spectrum possesses an exceptional point at momentum $k = 0$ (and $\xi_1 = \xi_2 = -1$).

\subsubsection{\chainfl{}: Transferred charge}

The Floquet-Bloch propagator from Eq.~\eqref{FloquetBloch} does not satisfy the prerequisite $[\hat U_{\mathsf{Fl}}^\dagger, \hat U_{\mathsf{Fl}}]$ of Eq.~\eqref{app:CBarNormal},
and we have to use Eq.~\eqref{app:charge_k} to obtain the transferred charge
\begin{equation}
\barC=2 s^4 \sinh 4\gamma \, + \, 2s^2 c^2 \sinh 2\gamma \;,
\end{equation}
reusing the previous abbreviations  $c\equiv \cos J$, $s\equiv \sin J$.
In the Hermitian case $\gamma=0$, we have $\bar C=0$.

The normalization factor 
\begin{equation}
\Xi^{1/2} = 1 + 2 s^2 \sinh^2 \gamma + 2 |s \sinh \gamma| \sqrt{1+s^2 \sinh^2 \gamma}
\end{equation}
is obtained from the dominant eigenvalue at $k=\pi$ (see Eq.~\eqref{eq:Fl_eigenvalues}).
Combining the expression for $\bar C$ and $\Xi$ we obtain an explicit expression for the normalized charge transfer $\bar C/\Xi$ plotted in Figs.~\eqref{fig:charge},~\eqref{fig:regularize}.

\subsubsection{\chainfl{}: Regularized dynamics}

Just as the static chains \chainst{}, \chainsp{}, the Floquet chain \chainfl{} has the trivial RD limit $\gamma = 0$ of a Hermitian chain, where $\bar C = 0$. 
In addition, there is the non-trivial RD limit $|\gamma| \to \infty$.

For large $|\gamma| \gg \gamma_c$, the
spectrum consists 
of
two loops $\xi_{1,2}(k) \sim - s^2 e^{\pm (\ii k - 2 \gamma) }$, separated by an imaginary gap $\ii \Gamma$ at $\Gamma=0$.
We have $\Xi \sim s^4 e^{4 |\gamma|}$. 
Being above the topological phase transition, the winding number is $W = \sgn \gamma \ne 0$.
The (normalized) transferred charge is  
\begin{equation}\label{barbarC_Fl}
\barC / \Xi = \frac{2 \sinh 4\gamma}{e^{4 |\gamma|}} + O(e^{-|\gamma|}) \;,
\end{equation}
which converges to $W$ in the limit $|\gamma| \to \infty$.
As expected from the general arguments, the transferred charge in the RD limit is quantized, equal to the winding number, and does not depend on the remaining model parameters (here, the hopping $J$).
This behavior is depicted in Fig.~\ref{fig:regularize}.

The propagator in the RD limit can be expressed in terms of the right ($S_+$) and left ($S_-$) shift operator
\begin{equation}\label{Shift}
S_+ = \sum_{n \in \mathbb Z} |2n+2 \rangle \langle 2n| \;,
 \quad
S_- = \sum_{n \in \mathbb Z} |2n-1 \rangle \langle 2n+1| \;.
\end{equation}
These operators shift wave packets exactly by two sites.
$S_+$ acts only on even sites (`filled' sites $\bullet$ in Fig.~\ref{fig:overview}, blue sites in Figs.~\ref{fig:charge},~\ref{fig:pump}), $S_-$ on odd sites (`open' or red sites $\circ$).

Shift operators are prototypical examples of charge pumps. Having non-zero winding number, they cannot be realized individually by continuous time propagation (which leaves the total winding number, summed over all states, invariant and equal to the initial value zero),  but have to appear in pairs.
Especially a simple shift operator $ \sum_{n \in \mathbb Z} |n+1 \rangle \langle n|$ cannot be realized by continuous time evolution.

In the RD limit of the Floquet chain, the normalized propagator $-U_\mathsf{Fl}/\Xi^{1/2}$ (with an additional minus sign) converges to $S_+$ for $\gamma \to \infty$, and to $S_-$ for $\gamma \to - \infty$. Here, the `missing' shift operator $S_-$ or $S_+$ is asymptotically suppressed by the strong damping $\sim e^{-4 |\gamma|}$ relative to the other operator $S_+$ or $S_-$ that survives in the RD limit.

\subsubsection{\chainfl{}: Perfect coupling}

All expressions for the chain \chainfl{} simplify in the case of perfect coupling $J = \pi/2$.
The name `perfect coupling' expresses the fact that amplitude is transferred perfectly between adjacent sites in each step, which is equivalent to the condition $c \equiv \cos J = 0$ in Eq.~\eqref{eq:real_space}.

At perfect coupling,
the spectrum of $\hat U_\mathsf{Fl}(k)$ consists of two perfectly circular loops $\xi_{1,2}(k) =- e^{\pm (\ii k - 2 \gamma) }$.
Note that $\gamma_c=0$: Away from the Hermitian case, the Floquet chain \chainfl{} at perfect coupling is always in a non-trivial topological phase.

The Floquet propagator in real space (see Eq.~\eqref{eq:real_space}) is  a weighted sum 
$U_\mathsf{Fl} = - e^{2 \gamma} S_+ - e^{-2 \gamma} S_-$ 
of the right and left shift operator $S_+$, $S_-$, even away from the RD limit.
In the RD limit, only one of the two shift operators survives, repeating the result given earlier for general $J$. 

The transferred charge is
\begin{equation}
 \barC / \Xi = \frac{2 \sinh 4 \gamma}{e^{4 |\gamma|}}
\end{equation}
for all $\gamma$ (cf. Eq.~\eqref{barbarC_Fl}).

\subsubsection{\chainfl{}: Disordered chain}

For Fig.~\ref{fig:pump} we add disorder to the hopping.
One can also include random detunings, and replace the two steps in Eq.~\eqref{eq:steps} with
\begin{subequations}
\begin{equation}
\begin{split}
 H_\mathsf{Fl}^{(1)} = \sum_{n \in \mathbb Z} \;\;\;  & J_n^{(a)} e^{\gamma} |2n+1\rangle \langle 2n| \\  + & J_n^{(b)} e^{-\gamma} |2n\rangle \langle 2n+1| 
 + \Delta_n^{(1)} \, |n\rangle \langle n| \; ,
 \end{split}
\end{equation}
\begin{equation}
\begin{split}
 H_\mathsf{Fl}^{(2)} = \sum_{n \in \mathbb Z}  \;\;\;  & J_n^{(c)} e^{\gamma} |2n\rangle \langle 2n-1| \\ + & J_n^{(d)} e^{-\gamma} |2n-1\rangle \langle 2n| + \Delta_n^{(2)} \, |n\rangle \langle n| \; ,
 \end{split}
\end{equation}
\end{subequations}
where the $J_n^{(a)}, ..., J_n^{(d)}$ are drawn independently and uniformly from the interval $[J(1-\delta),J(1+\delta)]$, with relative disorder strength $\delta$ just as in Sec.~\ref{sec:RobustTransport},
and the $\Delta_n^{(1)}, \Delta_n^{(2)}$ from  $[-\Delta, \Delta ]$.

The RD limit $|\gamma| \to \infty$ of this disordered Floquet chain can be analyzed completely
through a purely algebraic but rather lengthy computation of the Floquet propagator that we cannot repeat here.
The central result is that also in the disordered chain directional hopping suppresses one direction of propagation in favor of the other.

The limit $\gamma \to \infty$ of the normalized propagator $-U_\mathsf{Fl}/\Xi^{1/2}$ (still $\Xi = s^4 e^{4|\gamma|}$) results in a disordered shift operator
\begin{equation}\label{DisorderedShift}
S_+ = \sum_{n \in \mathbb Z} \zeta_n |2n+2 \rangle \langle 2n| \;,
\end{equation}
where the $\zeta_n$ are random variables of the same magnitude ($\zeta_n \approx 1$ for small $\delta$).
For $\delta = 0$, we recover the result $\zeta_n = 1$ of the ordered Floquet chain (see Eq.~\eqref{Shift}).
Analogously, a left shift is obtained for $\gamma \to -\infty$.

The action of the disorderd shift operator on wave packets is seen in Fig.~\ref{fig:pump}, panel \textsf{(B)},
which essentially shows $(S_+)^{n_p} |\psi\rangle$ for an initial Gaussian wave packet on $\approx 5$ sites.
The amplitude at individual sites depends on the $\zeta_n$, and thus is random,
but the entire wave packet propagates without spreading.

For a slightly simplified disordered Floquet chain even the transferred charge can be calculated explicitly, using Eqs.~\eqref{barcdis},~\eqref{xidis} derived from wave packet propagation.
To simplify, we drop the detunings ($\Delta_n^{(1,2)} \equiv 0$), and use equal disordered hopping ($J_n^{(a)} =  J_n^{(b)} \equiv J_n^{(1)}$, $J_n^{(c)} =  J_n^{(d)} \equiv J_n^{(2)}$) in each step.
Then, the wave functions $|\Psi_i\rangle = U_\mathsf{Fl} |i\rangle$ are given by
\begin{equation}
\begin{split}
|\Psi_{2i}\rangle = -\ii   \cos J_i^{(1)} \sin J_i^{(2)} \, e^{-\gamma} & |2i-1\rangle \\
+  \cos J_i^{(1)} \cos J_i^{(2)} \;\; & |2i\rangle \\
 -\ii \sin J_i^{(1)} \cos J_{i+1}^{(2)} \, e^\gamma & |2i+1\rangle \\
-  \sin J_i^{(1)} \sin J_{i+1}^{(2)} \, e^{2 \gamma} & |2i+2\rangle \;, \\[5pt]
|\Psi_{2i+1}\rangle =  - \sin J_i^{(1)} \sin J_i^{(2)} e^{-2\gamma} & |2i-1\rangle \\
- \ii \sin J_i^{(1)} \cos J_i^{(2)} e^{-\gamma} & |2i\rangle \\
  + \cos J_i^{(1)} \cos J_{i+1}^{(2)} & |2i+1\rangle \\
- \ii \cos J_i^{(1)} \sin J_{i+1}^{(2)} e^{\gamma} & |2i+2\rangle \;.
\end{split}
\end{equation}
We thus have
\begin{equation}
\begin{split}
 \langle \Psi_{2i} | \Psi_{2i} \rangle = &\cos^2 J_i^{(1)} \sin^2 J_i^{(2)}  e^{-2 \gamma} \\
&+ \cos^2 J_i^{(1)} \cos^2 J_i^{(2)} \\
& + \sin^2 J_i^{(1)} \cos^2 J_{i+1}^{(2)} e^{2 \gamma} \\
&+  \sin^2 J_i^{(1)} \sin^2 J_{i+1}^{(2)} e^{4 \gamma} \;, \\[5pt]
\langle \Psi_{2i+1} |\Psi_{2i+1}\rangle = &  \sin^2 J_i^{(1)} \sin^2 J_i^{(2)} e^{-4\gamma} \\
&+ \sin^2 J_i^{(1)} \cos^2 J_i^{(2)} e^{-2\gamma} \\
&  + \cos^2 J_i^{(1)} \cos^2 J_{i+1}^{(2)}\\
&+ \cos^2 J_i^{(1)} \sin^2 J_{i+1}^{(2)} e^{2\gamma} \;,
\end{split}
\end{equation}
and
\begin{equation}
\begin{split}
& \langle \Psi_{2i} |\hat x - 2 i | \Psi_{2i} \rangle =  - \cos^2 J_i^{(1)} \sin^2 J_i^{(2)}  e^{-2 \gamma} \\
 & \qquad + \sin^2 J_i^{(1)} \cos^2 J_{i+1}^{(2)} e^{2 \gamma} 
+ 2 \sin^2 J_i^{(1)} \sin^2 J_{i+1}^{(2)} e^{4 \gamma} \;, \\[5pt]
& \langle \Psi_{2i+1} |\hat x - (2 i+1) | \Psi_{2i+1} \rangle =  - 2 \sin^2 J_i^{(1)} \sin^2 J_i^{(2)} e^{-4\gamma} \\
& \qquad - \sin^2 J_i^{(1)} \cos^2 J_i^{(2)} e^{-2\gamma} 
+ \cos^2 J_i^{(1)} \sin^2 J_{i+1}^{(2)} e^{2\gamma} \;.
\end{split}
\end{equation}
To proceed, we require the averages of the $\cos^2, \sin^2$ terms over the uniform probability distribution of the $J_n^{(1)}, J_n^{(2)}$,
\begin{equation}\label{cossinave}
\begin{split}
\overline{c^2} &\equiv \overline{\cos^2 J_i} = \frac12 (1 - \zeta) + \zeta \cos^2 J \;, \\[5pt]
\overline{s^2} &\equiv \overline{\sin^2 J_i} = \frac12  (1 - \zeta) + \zeta \sin^2 J \;,
\end{split}
\end{equation}
where $\zeta = \sinc (2 J \delta)$ with the $\sinc$-function (with $\sinc x = (\sin x)/x$ for $x \ne 0$).

Therefore, the quantities in Eqs.~\eqref{barcdis},~\eqref{xidis} are
\begin{equation}\label{barcdis_fl}
\begin{split}
\barC^\mathsf{(wp)}  &= \frac{1}{2} \big(\overline{\langle \Psi_{2i} |\hat x - 2 i | \Psi_{2i} \rangle} + \overline{ \langle \Psi_{2i+1} |\hat x - (2 i+1) | \Psi_{2i+1} \rangle} \big) \\
&=  2 (\overline{s^2})^2 \sinh 4\gamma + 2 (\overline{s^2}) (\overline{c^2}) \sinh 2\gamma 
\end{split}
\end{equation}
and, using the prefactor $\eta=2 = L/N_\Gamma$ for the chain \chainfl{} (see the discussion after Eq.~\eqref{xidis}),
\begin{equation}\label{xidis_fl}
\begin{split}
\bar \Xi^\mathsf{(wp)}  &=\eta \frac{1}{2} \big(
\overline{\langle \Psi_{2i} | \Psi_{2i} \rangle} + \overline{\langle \Psi_{2i+1} |\Psi_{2i+1}\rangle}
 \big) \\
&= 2 (\overline{s^2})^2 \cosh 4\gamma  + 4 (\overline{s^2}) (\overline{c^2})  \cosh 2\gamma + 2 (\overline{c^2})^2 \;.
\end{split}
\end{equation}
The averages from Eq.~\eqref{cossinave} could now be inserted,
to obtain the normalized transferred charge $\barC^\mathsf{(wp)} / \bar \Xi^\mathsf{(wp)}$ of a disordered Floquet chain.
In the RD limit $|\gamma| \to \infty$, only the first term in the numerator and denominator survives, and we recover the result
$\barC^\mathsf{(wp)} / \bar \Xi^\mathsf{(wp)} = \sgn \gamma$ also with disorder.
For $\gamma \to \infty$, the parameters of the disordered shift operator from Eq.~\eqref{DisorderedShift} are given by $\zeta_n =  \sin J_n^{(1)} \sin J_n^{(2)}/\sin^2 J$. For $\delta=0$ we recover the result $\zeta_n \equiv 1$ of the ordered chain \chainfl{}.

\section{The transferred charge}
\label{app:TransferredCharge}

The transferred charge $C(n)$ measures the net amount of charge moving from the left part (sites $i<n$) to the right part (sites $i \ge n$) of a chain.

A wave function initially localized at site $|i\rangle$ evolves into $|\Psi_i\rangle = U |i \rangle$,
where the propagator $U$ is given in real space, e.g.,
as the Floquet propagator $U(T)$.
For a wave function starting in the left part 
of the chain, the amount of charge transferred to the right 
is $\sum_{j \ge n} |\langle j|\Psi_i\rangle|^2$. 
For a wave function starting in the right part 
of the chain, the amount of charge transferred to the left 
is $\sum_{j < n} |\langle j|\Psi_i\rangle|^2$, and has to be counted with a minus sign.
Therefore, the contribution of $|\Psi_i\rangle$ to $C(n)$ is
\begin{equation}\label{CnWavePacket}
C(n)|_i = \begin{cases}
\phantom- \sum_{j \ge n} |\langle j|\Psi_i\rangle|^2 \quad & \text{ for } i < n \;, \\[0.5ex]
- \sum_{j < n} |\langle j|\Psi_i\rangle|^2 \quad & \text{ for } i \ge n \;. \\
 \end{cases}
\end{equation}
Summing over all initial sites gives the transferred charge
\begin{equation}\label{app:Cn}
\begin{split}
C(n) &= \sum_{i \in \mathbb Z} C(n)|_i = \sum_{\substack{i < n \\ j \ge n}} |\langle j |\Psi_i\rangle|^2 - \sum_{\substack{i \ge n \\ j < n}} |\langle j |\Psi_i\rangle|^2  \\
&=\sum_{i \in \mathbb Z} \langle i | U^\dagger P_n U (1-P_n) - U^\dagger (1-P_n) U P_n  | i \rangle  \\
&= \sum_{i \in \mathbb Z} \langle i | U^\dagger P_n U - U^\dagger U P_n |i \rangle \\
&= \mathrm{tr}_\mathbb{Z} \, (U^\dagger [P_n,U]) \;,
\end{split}
\end{equation}
where the projection $P_n$ gives 
$P_n |i\rangle = \Theta (i-n) |i \rangle$ with
$\Theta(x)=1$ for $x \ge 0$, $\Theta(x) = 0$ for $x < 0$.
The trace is $\mathrm{tr}_\mathbb{Z} A = \sum_{i \in \mathbb Z} \langle i | A | i \rangle$.
This gives Eq.~\eqref{charge} in Sec.~\ref{sec:transport}.

Eq.~\eqref{app:Cn} generalizes the expressions for the Hermitian case (see, e.g., Eq.~(3.3) in Ref.~\cite{Graf2018}) to the non-Hermitian setting.
Note that the operator $U^\dagger [P_n,U]$ in the trace is not Hermitian, but 
$C(n) \in \mathbb R$ follows from the first line.

Similarly, we find the charge accumulated at site $n$ as
\begin{equation}
\begin{split}
c(n) &=\sum_{\substack{i \in \mathbb Z \\ i \ne n}} |\langle n|\Psi_i \rangle|^2 - \sum_{\substack{i \in \mathbb Z \\ i \ne n}} |\langle i|\Psi_n\rangle|^2 \\
&= \langle n| [U,U^\dagger] |n \rangle \;,
\end{split}
\end{equation}
the difference of charge moving to site $n$ and charge moving away from site $n$.

The accumulated charge gives the difference between $C(n)$ at different sites. We have, for $n \ge m$,
\begin{equation}\label{dummy}
\begin{split}
 C(m)& - C(n) = \sum_{i \in \mathbb Z} C(m)|_i - C(n)_i \\
 =&\phantom-  \sum_{\substack{i < m \\ j \ge m}} |\langle j|\Psi_i \rangle|^2
 -  \sum_{\substack{i \ge m \\ j  < m}} |\langle j|\Psi_i \rangle|^2 \\
& - \sum_{\substack{i < n \\ j \ge n}} |\langle j|\Psi_i \rangle|^2
 + \sum_{\substack{i \ge n \\ j < n}} |\langle j|\Psi_i \rangle|^2 \\
 =& \sum_{l=m}^{n-1} c(l) \;,
 \end{split}
\end{equation}
which expresses the conservation of charge, even in the non-Hermitian setting.
Hermiticity or regularized dynamics (RD), where $[U,U^\dagger]=0$, implies $c(n) = 0$ and thus $C(m) = C(n)$.

\subsection{The translationally invariant case}

Assume that $H$, and thus also $U$, is invariant under translations by $L$ sites, that is $U_{m+L,n+L} = U_{mn}$ for the matrix elements $U_{mn} = \langle m |U |n \rangle$ of the propagator.
Translational invariance implies 
$C(n)|_i = C(n+L)|_{i+L}$, in the notation of Eq.~\eqref{CnWavePacket}.

The charge $\barC$, averaged over a unit cell, is
\begin{equation}\label{app:barC}
\begin{split}
\bar C &= \frac{1}{L} \sum_{n=0}^{L-1} C(n) =\frac{1}{L} \sum_{n=0}^{L-1} \sum_{i \in \mathbb Z} C(n)|_i \\
&=\frac{1}{L} \sum_{n=0}^{L-1} \sum_{i=0}^{L-1} \sum_{m \in \mathbb Z} C(n)|_{i+mL} \\
&=\frac{1}{L} \sum_{n=0}^{L-1} \sum_{i=0}^{L-1} \sum_{m \in \mathbb Z} C(n-mL)|_{i} \\
&=\frac{1}{L} \sum_{i=0}^{L-1} \sum_{n \in \mathbb Z} C(n)|_i = \frac{1}{L} \sum_{i=0}^{L-1} \barC|_i\;,
\end{split}
\end{equation}
where we replace $i \to i + mL$ in line two, $n - mL \to n$ in line four, and use translational invariance in line three.
In the last line, we introduce the abbreviation 
\begin{equation}\label{barcidef}
\bar C |_i  = \, \sum_{n \in \mathbb Z} C(n) |_i \;.
\end{equation}
Note that translational invariance has allowed us to exchange the summation $0 \le n < L, i \in \mathbb Z$ with the summation $n \in \mathbb Z, 0 \le i < L$.

Inserting Eq.~\eqref{CnWavePacket} into Eq.~\eqref{barcidef} and rearranging summations we get
\begin{equation}\label{barcix}
\begin{split}
\bar C |_i & = \sum_{n > i} \sum_{j \ge n} \langle j| \Psi_i\rangle|^2 - \sum_{n \le i } \sum_{j < n} \langle j| \Psi_i\rangle|^2 \\
& =\sum_{j > i} (j-i) \langle j| \Psi_i\rangle|^2 - \sum_{j < i} (i-j) \langle j| \Psi_i  \rangle|^2 \\
& = \sum_{j \in \mathbb Z} (j-i) \, |\langle j|\Psi_i\rangle|^2  \, = \, \langle \Psi_i | \hat x - i | \Psi_i \rangle\;,
 \end{split}
\end{equation}
where $\hat x = \sum_{j \in \mathbb Z} j |j\rangle\langle j|$ is the position operator.
In other words, $\bar C|_i$ is the propagation distance of a wave packet initially localized at site $i$.
Therefore,  $\barC$ is the average propagation distance.

Note that in the non-Hermitian setting the norm of a wave function is not conserved, such that the propagation distance in Eq.~\eqref{barcix} is weighted by different $\langle \Psi_i| \Psi_i \rangle$.

\subsection{Momentum space expressions}
\label{app:MomentumSpace}

We define the Fourier transform as
\begin{equation}\label{eq:Fourier1}
\hat U_{mn}(k) = \sum_{l \in \mathbb Z} e^{-\ii (k/L)(m + l L - n)} \, U_{m+ l L, n} \;,
\end{equation}
which is an $L \times L$ matrix with indices $m, n \in \mathbb L := \{0, \dots, L-1 \}$, parametrized by momentum $k$. 
The inverse Fourier transform is
\begin{equation}\label{app:Fourier_trafo1}
 U_{m n} = \frac{1}{2\pi} \int_{-\pi}^\pi   \,  e^{\ii (k/L) \left( m - n  \right)} \, \hat U_{[m]_L, [n]_L}(k) \, \mathrm d k \;,
\end{equation}
where $[m]_L, [n]_L \in \mathbb L$ denotes the remainder after division by $L$.

A standard Fourier computation with Eq.~\eqref{app:Fourier_trafo1} gives
\begin{multline}
 \langle n |U^\dagger (\hat x - n) U | n \rangle = \\ \frac{\ii \mk L}{2\pi} \sum_{m=0}^{L-1}  \int_{-\pi}^\pi \hat U^\dagger_{[n]_L,m}(k) \, \partial_k  \hat U_{m,[n]_L}(k) \, \mathrm d k
 \end{multline}
for the interplay between the Fourier transform and the position operator $\hat x$.

Using this expression together with Eqs.~\eqref{app:barC},~\eqref{barcix} immediately gives
\begin{equation}\label{app:charge_k}
\bar C = \frac{\ii}{2\pi} \int_{-\pi}^{\pi}  \, \tr_\mathbb{L} \left(\hat U^\dagger(k) \, \partial_k \hat U(k)\right) \, \mathrm{d}k \;.
\end{equation}
Here, $\tr_\mathbb{L} A = \sum_{n=0}^{L-1} A_{nn}$ sums over the indices of the Fourier transform, which correspond to the $L$ sites of a unit cell.
This is the expression given in Sec.~\ref{sec:transport}.

Note that the Fourier transform in Eq.~\eqref{eq:Fourier1} is $2 \pi$-periodic only up to phase factors.
We have $\hat U(k+2\pi)=G(2 \pi) \hat U(k) G^{\dagger}(2\pi)$,
where $G(k)$ is a diagonal unitary matrix with entries $G_{nn}=\exp(-\ii k n/L)$.
By this relation, the spectrum of $\hat U(k)$ is $2 \pi$-periodic. 
Only $2 \pi$-periodicity of the spectrum, not of the Fourier transform, is required in the present derivations.

The alternative definition
\begin{equation}\label{eq:Fourier2}
\hat U_{mn}^\mathsf{(alt)}(k) = \sum_{l \in \mathbb Z} e^{-\ii k l} \, U_{m+ l L, n} \;,
\end{equation}
where the factor $e^{-\ii k l}$ is constant within a unit cell,
is $2\pi$-periodic.
The two Fourier transforms~\eqref{eq:Fourier1},~\eqref{eq:Fourier2} are connected by the unitary transformation $\hat U(k)=G(k) \hat U^\mathsf{(alt)}(k) G(k)^{\dagger}$,
such that the spectrum of the Floquet-Bloch propagator is independent of the choice of the Fourier transform.
Use of $\hat U_{mn}^\mathsf{(alt)}(k)$ would require replacing Eq.~\eqref{app:charge_k} by a more complicated expression that includes the matrix $G(k)$. This is why we prefer to work with $\hat U_{mn}(k) $.

We stress that, in contrast to the Hermitian case, Eq.~\eqref{app:charge_k} is \emph{not} invariant under $k$-dependent unitary transformations.
If we replace $\hat U(k)$ by $Q(k) \hat U(k) Q(k)^\dagger$, Eq.~\eqref{app:charge_k} changes into
\begin{equation}\label{app:charge_k_nogood}
\begin{split}
\bar C =  \frac{\ii}{2\pi} \int_{-\pi}^{\pi}  \, \tr_\mathbb{L} \Big( & \hat U^\dagger(k) \partial_k \hat U(k) \\
& + [\hat U(k), \hat U(k)^\dagger] Q(k)^\dagger \partial_k Q(k) \Big)
\end{split}
\end{equation}
(we have used $(\partial_k Q(k)^\dagger ) Q(k)= - Q(k)^\dagger \partial_k Q(k)$ for unitary $Q(k)$).
Through the additional term in the second line, $\bar C$ depends explicitly on the eigenvectors of $\hat U(k)$.
Therefore, we cannot diagonalize $\hat U(k)$ and express Eq.~\eqref{app:charge_k} entirely in terms of its eigenvalues. This complication is intrinsic to the non-Hermitian setting.

If, however, $[\hat U(k), \hat U^\dagger(k)]=0$, the additional term drops out of Eq.~\eqref{app:charge_k_nogood}.
This condition is known from linear algebra~\cite{GolubVanLoan}, where it defines a normal matrix.
Under this condition we can diagonalize the propagator with a unitary transformation as in the Hermitian case, write $\hat U(k) = Q(k) D(k) Q(k)^\dagger$ with a diagonal matrix $D(k)$ that contains the eigenvalues $\xi_1(k), \dots, \xi_L(k)$ of $\hat U(k)$,
and arrive at
\begin{equation}\label{app:CBarNormal}
\bar  C =
        \frac{\ii}{2\pi} \sum_{m=1}^L  \int_{-\pi}^{\pi}   \xi_m(k)^* \, \partial_k \xi_m(k) \, \mathrm{d}k \;.
\end{equation}
This is Eq.~\eqref{IbarNormal} in Sec.~\ref{sec:transport}.
It does not depend on the choice of the Fourier transform.

Since $\bar C \in \mathbb R$, we can rewrite this equation as
\begin{equation}
\bar  C =
    - \frac{1}{2\pi} \sum_{m=1}^L  \int_{-\pi}^{\pi}  \Im \big( \xi_m(k)^* \, \partial_k \xi_m(k) \big) \, \mathrm{d}k \;,
\end{equation}
which suggests the geometric interpretation that the transferred charge $\bar C$
 is the area in the $\xi$-plane enclosed by the eigenvalue loops $k \mapsto \xi_m(k)$.
The sign of $\bar C$ is such that a clockwise loop gives a positive contribution.

\subsection{Regularized dynamics}

The transferred charge $\barC$ is not invariant under (imaginary) shifts of the Hamiltonian $H \mapsto H + s$,
where $U(T) \mapsto e^{T s} U(T)$, hence  $\bar C \mapsto \bar C + e^{2 \mk T \Im s}$. 
Therefore, it is useful to consider the normalized transferred charge $\barC / \Xi$,
with an appropriate normalization factor $\Xi$.
We set $\Xi = \max_\xi |\xi|^2$ to the absolute square of the eigenvalues $\xi$ of $U$, i.e., the square of the spectral radius
(normalization for disordered chains is addressed in the next subsection).
In this way, the dominant eigenvalue of the normalized propagator $U/\Xi^{1/2}$ has modulus one.
This is the normalization chosen in Sec.~\ref{sec:RegularizedDynamics} to define the RD limit, where the `surviving' loops should have zero gain or loss.

RD as defined in Sec.~\ref{sec:RegularizedDynamics} requires that the eigenvectors of $U$ (or $\hat U(k)$) are orthogonal and the eigenvalues fulfill $|\xi|^2 = |\xi|$, i.e., have modulus zero or one. These conditions are equivalent to the statement that $U$ is a normal operator (orthogonal eigenvectors) and a partial isometry ($|\xi| \in \{0,1\}$),
that is, $P = U^\dagger U = U U^\dagger$ and $P$ is a projection ($P^2 = P$).
This generalizes the Hermitian case, where $U$ is unitary and $P = \mathbbm 1$.
Note that the statement $(U U^\dagger)^2 = U U^\dagger$ is equivalent to $U^\dagger U U^\dagger = U^\dagger$.

How the RD conditions allow one to relate transport and topology is explored in Sec.~\ref{sec:RegularizedDynamics}.
The RD conditions can also be motivated if we consider the transferred charge as a function $\barC(t)$ of time, or as a function $\barC(n_p)$ of periods $n_p$ in a Floquet chain,
and ask for the persistent current given by the limit $\lim_{t \to \infty} \barC(t)/t$ or $\lim_{n_p \to \infty} \barC(n_p)/n_p$.
For a (topological) charge pump the persistent current should be non-zero.

This leads us to ask for the conditions under which
\begin{equation}\label{CMultiple}
\barC (n_p) = n_p \, \barC
\end{equation}
holds.
This equation is generally not true.
For example, we have
\begin{equation}
\begin{split}
\bar C(2) &= \frac{\ii}{2\pi} \int_{-\pi}^{\pi}  \, \tr_\mathbb{L} \left(\hat {U^\dagger}^2 \, \partial_k \big(\hat U^2\big) \right) \, \mathrm{d}k  \\
&= \frac{\ii}{2\pi} \int_{-\pi}^{\pi}  \, \tr_\mathbb{L} \left( \big(\hat U \hat {U^\dagger}^2  + \hat {U^\dagger}^2 \hat U  \big) \, \partial_k \hat U \right) \, \mathrm{d}k
\end{split}
\end{equation}
from Eq.~\eqref{app:charge_k}, if we replace $\hat U \equiv  U(T,k)$ by $U(2 T,k ) = \hat U^2$. 
To make progress towards Eq.~\eqref{CMultiple} we have to relate the operator expression $\hat U \hat {U^\dagger}^2  + \hat {U^\dagger}^2 \hat U$ to $2 \hat U^\dagger$.
Without aiming for a mathematically strict statement we note that if $\hat U^\dagger \hat U = \hat U \hat U^\dagger$ we have $\hat U \hat {U^\dagger}^2  + \hat {U^\dagger}^2 \hat U = 2 \hat U^\dagger \hat U  \hat U^\dagger$,
and we obtain the desired relation if additionally $\hat U^\dagger \hat U  \hat U^\dagger = \hat U^\dagger$.
These are precisely the RD conditions formulated above.

That, conversely, RD implies Eq.~\eqref{CMultiple} is obvious with Eq.~\eqref{app:CBarNormal} (which can be used here),
if $\xi_m(k)$ is replaced by $\xi_m(k)^{n_p}$ and one uses $|\xi_m(k)| = 1$ for the eigenvalue loops that contribute in this equation.

Therefore, in the RD limit, a non-zero charge transfer $\bar C \ne 0$ corresponds to a non-zero persistent current $\lim_{n_p \to \infty} \barC(n_p)/n_p = \bar C$, which is given by the relation $\bar C = W(\Gamma)$ between transport and topology.

\subsection{Measuring the transferred charge by wave packet propagation}

The transferred charge $C(n)$ or $\bar C$ can be computed or measured also with real-space wave packet propagation.
The principal procedure is to prepare a wave packet $|\psi(0)\rangle = |i\rangle$ at a single site $i$ of the chain, let it evolve in time to $|\Psi_i\rangle = |\psi(T)\rangle = U |\psi(0)\rangle$,
and then determine the quantity $C(n)|_i$ in Eq.~\eqref{CnWavePacket} or the quantity $\barC_i$ in Eq.~\eqref{barcix}.
For $C(n)$ the contribution from wave packets for all sites of the chain has to be summed,
but $C(n)|_i$ will be negligible for $|i-n| \gg 1$.
Owing to translational invariance, the averaged charge $\barC$ can be obtained exactly from the propagation of only $L$ individual wave packets, prepared at the sites $i \in \{0, \dots, L-1 \}$ of the unit cell.
Experimentally, this procedure requires preparation of a wave packet at a single site, and measurement of the weight of the propagated wave packet at multiple sites.
This is possible, e.g., in experiments using photonic waveguide lattices~\cite{SzameitJPB,RevModPhys.91.015006}.

In other situations one may wish for a more relaxed approach that does not require preparation of a wave packet at a single site.
A practical way is to measure the persistent current of a Floquet chain close to the RD limit:
let a wave packet $|\psi\rangle$ propagate over $n_p \ge 1$ periods to $|\Psi\rangle = U^{n_p} |\psi\rangle$,
measure the propagation distance $\Delta x = (\langle \Psi |\hat x|\Psi \rangle - \langle \psi|\hat x|\psi\rangle)/\langle \Psi|\Psi\rangle$, appropriately normalized by $\langle \Psi|\Psi\rangle$, and approximate the transferred charge by $\barC \approx \Delta x/n_p$.
If the width of the initial wave packet is small compared to the propagation distance, but still large enough to average over multiple sites of the chain, this will give a good approximation of $\bar C$ for a (topological) charge pump. Sampling over multiple wave packets is possible.

Wave packet propagation remains applicable in a disordered chain, where the momentum space expressions~\eqref{app:charge_k},~\eqref{app:CBarNormal} cannot be used.
We can still define the transferred charge $\barC$ as an average
\begin{equation}\label{barcdis}
\barC^\mathsf{(wp)} = \lim_{n \to \infty} \frac{1}{2n+1} \sum_{i=-n}^n \barC_i \;,
\end{equation}
now over the entire chain instead of only one unit cell.

In this context, one has to reconsider the meaning of normalization, since the normalization factor $\Xi$ can no longer be easily set to the square of the spectral radius of the (Floquet-Bloch) propagator.
Instead, we can also consider an average
\begin{equation}\label{xidis}
\bar \Xi^\mathsf{(wp)} = \eta \lim_{n \to \infty} \frac{1}{2n+1} \sum_{i=-n}^n \langle \Psi_i | \Psi_i \rangle \;,
\end{equation}
over the norm of wave packets $|\Psi_i\rangle = U|i\rangle$ starting from all sites of the chain.
This expression for $\bar \Xi^\mathsf{(wp)}$ contains a prefactor $\eta$ that has to be determined from comparison with the previous normalization for a chain without disorder.
Reverting to momentum space, we find
\begin{equation}
 \lim_{n \to \infty} \frac{1}{2n+1} \sum_{i=-n}^n \langle \Psi_i | \Psi_i \rangle = \frac{1}{2\pi L} \sum_{m=1}^L  \int_{-\pi}^{\pi}   |\xi_m(k)|^2 \, \mathrm{d}k
\end{equation}
for a unit cell of $L$ sites.
In the RD limit, we thus have $\bar \Xi^\mathsf{(wp)} = \eta (N_\Gamma / L) \mk \Xi$ in comparison to the previous factor $\Xi = \max_\xi |\xi|^2$,
where $N_\Gamma$ is the number of `surviving' eigenvalue loops above the imaginary gap.
Therefore, we set $\eta = L/N_\Gamma$.
For the chains \chainsp, \chainfl,  with $L =2, N_\Gamma=1$, this means $\eta = 2$.
In this way, the expressions for a disordered chain agree with the previous expressions for a chain without disorder at least in the RD limit. 
A difference persists away from this limit, and other ways of normalization are certainly possible.

Eqs.~\eqref{barcdis},~\eqref{xidis} are used to compute the transferred charge 
for the inset of Fig.~\ref{fig:pump}. Explicit expressions are given in Eqs.~\eqref{barcdis_fl},~\eqref{xidis_fl}.

\section{Regularization of the propagator}
\label{app:Regularization}

In this section all matrices depend on momentum $k$,
and we take it for granted that matrix functions $k \mapsto \hat U(k)$ etc. are at least continuous.
We also demand $2 \pi$-periodicity of such functions, but only up to phase factors, similar to the behavior of the Fourier transform in Eq.~\eqref{eq:Fourier1}.
In the following equations, we occasionally drop the $k$-dependence to allow for simpler notation.

Regularization starts with a Floquet-Bloch propagator $\hat U(k)$ with imaginary gap $\ii \Gamma$. 
For a unit cell of $L$ sites, $\hat U(k)$ is an $L \times L$ matrix.
If the eigenspace to eigenvalues $|\xi_m(k)| > e^\Gamma$ outside of the imaginary gap has dimension $l$, with $1 \le l \le L$, the eigenspace to eigenvalues $|\xi_m(k)| < e^\Gamma$ inside of the imaginary gap has dimension $L-l$.
Note that $l$ does not depend on $k$.

We first consider the case that the eigenvectors of the propagator are already orthogonal before regularization, that is $[\hat U, \hat U^\dagger]=0$ (this is the case, e.g., for the static chains \chainst{}, \chainsp{}, but not for the Floquet chain \chainfl{}).
Then, regularization deforms only the eigenvalues of $\hat U$.

To implement the deformation, diagonalize $\hat U(k)$ with a unitary transformation $Q(k)$,
such that $\hat U(k) = Q(k) D(k) Q(k)^\dagger$ with a diagonal matrix $D(k)$ that contains the eigenvalues $\xi_1(k), \dots, \xi_L(k)$ of $\hat U(k)$.
The deformation of the propagator is given by
\begin{equation}
 \breve U(s) = Q  \, f_\Gamma(D,s) \, Q^\dagger \;,
\end{equation}
with the function
\begin{equation}
 f_\Gamma(z,s) = \begin{cases}
  (1-s) z \quad & \text{ if } |z| < e^\Gamma \;, \\[5pt]
  \Big(1 -s + \dfrac{s}{|z|} \Big) z & \text{ if } |z| > e^\Gamma \;.
 \end{cases}
\end{equation}
We have $f_\Gamma(z,0) = z$, and $f_\Gamma(z,1) = 0$ for $|z| < e^\Gamma$ but $f_\Gamma(z,1) = z/|z|$ for $|z| > e^\Gamma$.
$ \breve U(s)$ does not depend on the choice of the transformation $Q$,
and we are justified to write directly $\breve U(s) = f_\Gamma(\hat U,s)$.

With this definition, 
$\breve U(0) = \hat U$,
while the eigenvalues of $f_\Gamma(D,1)$, hence of $\breve U(1)$, have modulus zero or one.
Furthermore, 
eigenvalues move radially towards the unit circle or the origin
such that an imaginary gap stays open during the deformation from $\breve U(0)$ to $\breve U(1)$.
This completes regularization in this situation.

In the general case $[\hat U, \hat U^\dagger] \ne 0$, also the eigenvectors of $\hat U$ must be deformed during regularization. One approach is to use the Schur decomposition~\cite{GolubVanLoan},
where we write the propagator as $\hat U(k) = Q(k) A(k) Q(k)^\dagger$ with unitary $Q(k)$ and triangular $A(k)$.
One reason to use the Schur decomposition is that, e.g., the spectral decomposition fails to exist at exceptional points.

Write $A(k) =D(k) + N(k)$, where the diagonal part $D(k)$ contains the eigenvalues of $\hat U(k)$, and $N(k)$ is strictly upper triangular.
Note that $N(k)$ is zero if and only if $[\hat U, \hat U^\dagger] = 0$.

Now we define the continuous deformation 
\begin{equation}\label{DeformUSchur}
 A(s) = f_\Gamma(D,s) + (1-s) N 
\end{equation}
for $0 \le s \le 1$.
This deformation still moves the eigenvalues on the diagonal of $A$ radially towards the unit circle or the origin, and sends the triangular part $N$ above the diagonal to zero such that $[A(1),A(1)^\dagger]=0$.
Reinserting into the Schur decomposition gives a continuous deformation of $\hat U(k)$ to a regularized propagator. 

The problem with using the Schur decomposition is that the deformation of $A(k)$ depends explicitly on $N$, and is not invariant under unitary transformations.
Conceptually, for $N \ne 0$ the deformation in Eq.~\eqref{DeformUSchur} involves a specific choice how the eigenvectors of $A(k)$ (equivalently, of $\hat U(k)$) are made orthogonal during regularization.
Since the Schur decomposition is not unique, it can happen that $N(0) \ne N(2 \pi)$.
Then, the Schur-based regularization can fail to preserve the $2 \pi$-periodicity of $\hat U(k)$, even if required only up to phase factors.

Incidentally, for the present examples with $l=L-l=1$, this problem does not arise. Here, the Schur decomposition is unique (up to phase factors) since the imaginary gap separates the one-dimensional eigenspaces for all $k$.

We shall briefly introduce an alternative approach to regularization that avoids the complications of the Schur decomposition.
Let $P_>(k)$ denote the projection onto the eigenspace to eigenvalues $|\xi(k)| > e^\Gamma$ outside of the imaginary gap, and $P_<(k)$ the projection onto the eigenspace to eigenvalues $|\xi(k)| <e^\Gamma$ inside of the imaginary gap.
Note that we cannot assume that the projections are orthogonal, unless $\hat U$ is normal.

Using the projections, split $\hat U(k)$ as
\begin{equation}
\begin{split}
 \hat U(k) &= P_> \hat U  P_> \,+\, P_< \hat U P_< \\
 & = \hat U_> \, + \, \hat  U_< \;.
 \end{split}
\end{equation}
We now deform $\hat U_>$ to a partial isometry, and let $\hat U_< \to 0$.

By assumption, $\hat U_>$ is a matrix with rank $l$. Therefore, it can be written
in the form
\begin{equation}\label{UgtrQB}
\hat U_> = Q B Q^\dagger \;,
\end{equation}
where $Q$ is a $L \times l$ matrix with orthogonal columns (such that $Q^\dagger Q = \mathbbm 1_l$), and $B$ an $l \times l$ matrix. The spectrum of $B$ is equal to the non-zero  spectrum of $\hat U_>$, in particular, $B$ has full rank $l$.

Note that here no problem arises since we do not assume a specific form of $B$ as we did in the Schur decomposition for the matrix $A$. The functions $k \mapsto Q(k), B(k)$ can always be chosen with periodicity $2 \pi$.

Now we use a polar decomposition~\cite{GolubVanLoan}
\begin{equation}
 B =  R S \;,
\end{equation}
with a unitary $l \times l$ matrix $R$ and a Hermitian and positive definite $l \times l$ matrix $S$ (the latter is guaranteed since $B$ has full rank $l$).
Using the polar decomposition, which is unique for full rank $B$, avoids the complications arising from the non-uniqueness of the Schur decomposition.

Since $S$ is positive definite, we can write $S =e^X$ with Hermitian $X$,
and define a continuous deformation
\begin{equation}\label{DeformUPolar}
 \breve U(s) = Q R e^{(1-s) X} Q^\dagger  \, + \, g(s) \hat U_< 
\end{equation}
of $\hat U$ with parameter $s \in [0,1]$.
Here, $g(s)$ is a (largely arbitrary) function with $g(s) > 0$, $g(0)=1$, and $g(1)=0$.

The map $s \mapsto \breve U(s)$ is a continuous deformation of  the propagator,
such that $\breve U(1) = Q R Q^\dagger$ is a regularized propagator. 
Since the smallest eigenvalue of the matrix $Q R e^{(1-s) X} Q^\dagger$ is bounded away from zero for $0 \le s \le 1$,
we can always achieve that an imaginary gap stays open during the deformation by letting $g(s)$ approach zero sufficiently fast.

Note that the deformation~\eqref{DeformUPolar} does not move eigenvalues along radial lines as the previous deformation~\eqref{DeformUSchur} did, unless $\hat U$ is normal and we can choose a diagonal $B$ in Eq.~\eqref{UgtrQB}.
We recall that, regarding topological properties, details of the regularization procedure are irrelevant.


\begin{thebibliography}{48}%
\makeatletter
\providecommand \@ifxundefined [1]{%
 \@ifx{#1\undefined}
}%
\providecommand \@ifnum [1]{%
 \ifnum #1\expandafter \@firstoftwo
 \else \expandafter \@secondoftwo
 \fi
}%
\providecommand \@ifx [1]{%
 \ifx #1\expandafter \@firstoftwo
 \else \expandafter \@secondoftwo
 \fi
}%
\providecommand \natexlab [1]{#1}%
\providecommand \enquote  [1]{``#1''}%
\providecommand \bibnamefont  [1]{#1}%
\providecommand \bibfnamefont [1]{#1}%
\providecommand \citenamefont [1]{#1}%
\providecommand \href@noop [0]{\@secondoftwo}%
\providecommand \href [0]{\begingroup \@sanitize@url \@href}%
\providecommand \@href[1]{\@@startlink{#1}\@@href}%
\providecommand \@@href[1]{\endgroup#1\@@endlink}%
\providecommand \@sanitize@url [0]{\catcode `\\12\catcode `\$12\catcode
  `\&12\catcode `\#12\catcode `\^12\catcode `\_12\catcode `\%12\relax}%
\providecommand \@@startlink[1]{}%
\providecommand \@@endlink[0]{}%
\providecommand \url  [0]{\begingroup\@sanitize@url \@url }%
\providecommand \@url [1]{\endgroup\@href {#1}{\urlprefix }}%
\providecommand \urlprefix  [0]{URL }%
\providecommand \Eprint [0]{\href }%
\providecommand \doibase [0]{http://dx.doi.org/}%
\providecommand \selectlanguage [0]{\@gobble}%
\providecommand \bibinfo  [0]{\@secondoftwo}%
\providecommand \bibfield  [0]{\@secondoftwo}%
\providecommand \translation [1]{[#1]}%
\providecommand \BibitemOpen [0]{}%
\providecommand \bibitemStop [0]{}%
\providecommand \bibitemNoStop [0]{.\EOS\space}%
\providecommand \EOS [0]{\spacefactor3000\relax}%
\providecommand \BibitemShut  [1]{\csname bibitem#1\endcsname}%
\let\auto@bib@innerbib\@empty
\bibitem [{\citenamefont {Klitzing}\ \emph {et~al.}(1980)\citenamefont
  {Klitzing}, \citenamefont {Dorda},\ and\ \citenamefont {Pepper}}]{Klitzing}%
  \BibitemOpen
  \bibfield  {author} {\bibinfo {author} {\bibfnamefont {K.~v.}\ \bibnamefont
  {Klitzing}}, \bibinfo {author} {\bibfnamefont {G.}~\bibnamefont {Dorda}}, \
  and\ \bibinfo {author} {\bibfnamefont {M.}~\bibnamefont {Pepper}},\ }\href
  {\doibase 10.1103/PhysRevLett.45.494} {\bibfield  {journal} {\bibinfo
  {journal} {Phys. Rev. Lett.}\ }\textbf {\bibinfo {volume} {45}},\ \bibinfo
  {pages} {494} (\bibinfo {year} {1980})}\BibitemShut {NoStop}%
\bibitem [{\citenamefont {Thouless}\ \emph {et~al.}(1982)\citenamefont
  {Thouless}, \citenamefont {Kohmoto}, \citenamefont {Nightingale},\ and\
  \citenamefont {den Nijs}}]{TKNN}%
  \BibitemOpen
  \bibfield  {author} {\bibinfo {author} {\bibfnamefont {D.~J.}\ \bibnamefont
  {Thouless}}, \bibinfo {author} {\bibfnamefont {M.}~\bibnamefont {Kohmoto}},
  \bibinfo {author} {\bibfnamefont {M.~P.}\ \bibnamefont {Nightingale}}, \ and\
  \bibinfo {author} {\bibfnamefont {M.}~\bibnamefont {den Nijs}},\ }\href
  {\doibase 10.1103/PhysRevLett.49.405} {\bibfield  {journal} {\bibinfo
  {journal} {Phys. Rev. Lett.}\ }\textbf {\bibinfo {volume} {49}},\ \bibinfo
  {pages} {405} (\bibinfo {year} {1982})}\BibitemShut {NoStop}%
\bibitem [{\citenamefont {Kane}\ and\ \citenamefont
  {Mele}(2005)}]{KaneMelePRL}%
  \BibitemOpen
  \bibfield  {author} {\bibinfo {author} {\bibfnamefont {C.~L.}\ \bibnamefont
  {Kane}}\ and\ \bibinfo {author} {\bibfnamefont {E.~J.}\ \bibnamefont
  {Mele}},\ }\href {\doibase 10.1103/PhysRevLett.95.146802} {\bibfield
  {journal} {\bibinfo  {journal} {Phys. Rev. Lett.}\ }\textbf {\bibinfo
  {volume} {95}},\ \bibinfo {pages} {146802} (\bibinfo {year}
  {2005})}\BibitemShut {NoStop}%
\bibitem [{\citenamefont {K{\"o}nig}\ \emph {et~al.}(2007)\citenamefont
  {K{\"o}nig}, \citenamefont {Wiedmann}, \citenamefont {Br{\"u}ne},
  \citenamefont {Roth}, \citenamefont {Buhmann}, \citenamefont {Molenkamp},
  \citenamefont {Qi},\ and\ \citenamefont {Zhang}}]{Konig2007}%
  \BibitemOpen
  \bibfield  {author} {\bibinfo {author} {\bibfnamefont {M.}~\bibnamefont
  {K{\"o}nig}}, \bibinfo {author} {\bibfnamefont {S.}~\bibnamefont {Wiedmann}},
  \bibinfo {author} {\bibfnamefont {C.}~\bibnamefont {Br{\"u}ne}}, \bibinfo
  {author} {\bibfnamefont {A.}~\bibnamefont {Roth}}, \bibinfo {author}
  {\bibfnamefont {H.}~\bibnamefont {Buhmann}}, \bibinfo {author} {\bibfnamefont
  {L.~W.}\ \bibnamefont {Molenkamp}}, \bibinfo {author} {\bibfnamefont {X.-L.}\
  \bibnamefont {Qi}}, \ and\ \bibinfo {author} {\bibfnamefont {S.-C.}\
  \bibnamefont {Zhang}},\ }\href {\doibase 10.1126/science.1148047} {\bibfield
  {journal} {\bibinfo  {journal} {Science}\ }\textbf {\bibinfo {volume}
  {318}},\ \bibinfo {pages} {766} (\bibinfo {year} {2007})}\BibitemShut
  {NoStop}%
\bibitem [{\citenamefont {Hasan}\ and\ \citenamefont
  {Kane}(2010)}]{HasanKane2010}%
  \BibitemOpen
  \bibfield  {author} {\bibinfo {author} {\bibfnamefont {M.~Z.}\ \bibnamefont
  {Hasan}}\ and\ \bibinfo {author} {\bibfnamefont {C.~L.}\ \bibnamefont
  {Kane}},\ }\href {\doibase 10.1103/RevModPhys.82.3045} {\bibfield  {journal}
  {\bibinfo  {journal} {Rev. Mod. Phys.}\ }\textbf {\bibinfo {volume} {82}},\
  \bibinfo {pages} {3045} (\bibinfo {year} {2010})}\BibitemShut {NoStop}%
\bibitem [{\citenamefont {Qi}\ and\ \citenamefont
  {Zhang}(2011)}]{RevModPhys.83.1057}%
  \BibitemOpen
  \bibfield  {author} {\bibinfo {author} {\bibfnamefont {X.-L.}\ \bibnamefont
  {Qi}}\ and\ \bibinfo {author} {\bibfnamefont {S.-C.}\ \bibnamefont {Zhang}},\
  }\href {\doibase 10.1103/RevModPhys.83.1057} {\bibfield  {journal} {\bibinfo
  {journal} {Rev. Mod. Phys.}\ }\textbf {\bibinfo {volume} {83}},\ \bibinfo
  {pages} {1057} (\bibinfo {year} {2011})}\BibitemShut {NoStop}%
\bibitem [{\citenamefont {Su}\ \emph {et~al.}(1979)\citenamefont {Su},
  \citenamefont {Schrieffer},\ and\ \citenamefont
  {Heeger}}]{PhysRevLett.42.1698}%
  \BibitemOpen
  \bibfield  {author} {\bibinfo {author} {\bibfnamefont {W.~P.}\ \bibnamefont
  {Su}}, \bibinfo {author} {\bibfnamefont {J.~R.}\ \bibnamefont {Schrieffer}},
  \ and\ \bibinfo {author} {\bibfnamefont {A.~J.}\ \bibnamefont {Heeger}},\
  }\href {\doibase 10.1103/PhysRevLett.42.1698} {\bibfield  {journal} {\bibinfo
   {journal} {Phys. Rev. Lett.}\ }\textbf {\bibinfo {volume} {42}},\ \bibinfo
  {pages} {1698} (\bibinfo {year} {1979})}\BibitemShut {NoStop}%
\bibitem [{\citenamefont {Kitaev}(2001)}]{Kitaev_2001}%
  \BibitemOpen
  \bibfield  {author} {\bibinfo {author} {\bibfnamefont {A.~Y.}\ \bibnamefont
  {Kitaev}},\ }\href {\doibase 10.1070/1063-7869/44/10s/s29} {\bibfield
  {journal} {\bibinfo  {journal} {Phys.-Usp.}\ }\textbf {\bibinfo {volume}
  {44}},\ \bibinfo {pages} {131} (\bibinfo {year} {2001})}\BibitemShut
  {NoStop}%
\bibitem [{\citenamefont {Thouless}(1983)}]{PhysRevB.27.6083}%
  \BibitemOpen
  \bibfield  {author} {\bibinfo {author} {\bibfnamefont {D.~J.}\ \bibnamefont
  {Thouless}},\ }\href {\doibase 10.1103/PhysRevB.27.6083} {\bibfield
  {journal} {\bibinfo  {journal} {Phys. Rev. B}\ }\textbf {\bibinfo {volume}
  {27}},\ \bibinfo {pages} {6083} (\bibinfo {year} {1983})}\BibitemShut
  {NoStop}%
\bibitem [{\citenamefont {Xiao}\ \emph {et~al.}(2010)\citenamefont {Xiao},
  \citenamefont {Chang},\ and\ \citenamefont {Niu}}]{RevModPhys.82.1959}%
  \BibitemOpen
  \bibfield  {author} {\bibinfo {author} {\bibfnamefont {D.}~\bibnamefont
  {Xiao}}, \bibinfo {author} {\bibfnamefont {M.-C.}\ \bibnamefont {Chang}}, \
  and\ \bibinfo {author} {\bibfnamefont {Q.}~\bibnamefont {Niu}},\ }\href
  {\doibase 10.1103/RevModPhys.82.1959} {\bibfield  {journal} {\bibinfo
  {journal} {Rev. Mod. Phys.}\ }\textbf {\bibinfo {volume} {82}},\ \bibinfo
  {pages} {1959} (\bibinfo {year} {2010})}\BibitemShut {NoStop}%
\bibitem [{\citenamefont {Privitera}\ \emph {et~al.}(2018)\citenamefont
  {Privitera}, \citenamefont {Russomanno}, \citenamefont {Citro},\ and\
  \citenamefont {Santoro}}]{PhysRevLett.120.106601}%
  \BibitemOpen
  \bibfield  {author} {\bibinfo {author} {\bibfnamefont {L.}~\bibnamefont
  {Privitera}}, \bibinfo {author} {\bibfnamefont {A.}~\bibnamefont
  {Russomanno}}, \bibinfo {author} {\bibfnamefont {R.}~\bibnamefont {Citro}}, \
  and\ \bibinfo {author} {\bibfnamefont {G.~E.}\ \bibnamefont {Santoro}},\
  }\href {\doibase 10.1103/PhysRevLett.120.106601} {\bibfield  {journal}
  {\bibinfo  {journal} {Phys. Rev. Lett.}\ }\textbf {\bibinfo {volume} {120}},\
  \bibinfo {pages} {106601} (\bibinfo {year} {2018})}\BibitemShut {NoStop}%
\bibitem [{\citenamefont {Shen}\ \emph {et~al.}(2018)\citenamefont {Shen},
  \citenamefont {Zhen},\ and\ \citenamefont {Fu}}]{PhysRevLett.120.146402}%
  \BibitemOpen
  \bibfield  {author} {\bibinfo {author} {\bibfnamefont {H.}~\bibnamefont
  {Shen}}, \bibinfo {author} {\bibfnamefont {B.}~\bibnamefont {Zhen}}, \ and\
  \bibinfo {author} {\bibfnamefont {L.}~\bibnamefont {Fu}},\ }\href
  {https://link.aps.org/doi/10.1103/PhysRevLett.120.146402} {\bibfield
  {journal} {\bibinfo  {journal} {Phys. Rev. Lett.}\ }\textbf {\bibinfo
  {volume} {120}},\ \bibinfo {pages} {146402} (\bibinfo {year}
  {2018})}\BibitemShut {NoStop}%
\bibitem [{\citenamefont {Lin}\ \emph {et~al.}(2011)\citenamefont {Lin},
  \citenamefont {Ramezani}, \citenamefont {Eichelkraut}, \citenamefont
  {Kottos}, \citenamefont {Cao},\ and\ \citenamefont
  {Christodoulides}}]{PhysRevLett.106.213901}%
  \BibitemOpen
  \bibfield  {author} {\bibinfo {author} {\bibfnamefont {Z.}~\bibnamefont
  {Lin}}, \bibinfo {author} {\bibfnamefont {H.}~\bibnamefont {Ramezani}},
  \bibinfo {author} {\bibfnamefont {T.}~\bibnamefont {Eichelkraut}}, \bibinfo
  {author} {\bibfnamefont {T.}~\bibnamefont {Kottos}}, \bibinfo {author}
  {\bibfnamefont {H.}~\bibnamefont {Cao}}, \ and\ \bibinfo {author}
  {\bibfnamefont {D.~N.}\ \bibnamefont {Christodoulides}},\ }\href
  {https://link.aps.org/doi/10.1103/PhysRevLett.106.213901} {\bibfield
  {journal} {\bibinfo  {journal} {Phys. Rev. Lett.}\ }\textbf {\bibinfo
  {volume} {106}},\ \bibinfo {pages} {213901} (\bibinfo {year}
  {2011})}\BibitemShut {NoStop}%
\bibitem [{\citenamefont {Harari}\ \emph {et~al.}(2018)\citenamefont {Harari},
  \citenamefont {Bandres}, \citenamefont {Lumer}, \citenamefont {Rechtsman},
  \citenamefont {Chong}, \citenamefont {Khajavikhan}, \citenamefont
  {Christodoulides},\ and\ \citenamefont {Segev}}]{Hararieaar4003}%
  \BibitemOpen
  \bibfield  {author} {\bibinfo {author} {\bibfnamefont {G.}~\bibnamefont
  {Harari}}, \bibinfo {author} {\bibfnamefont {M.~A.}\ \bibnamefont {Bandres}},
  \bibinfo {author} {\bibfnamefont {Y.}~\bibnamefont {Lumer}}, \bibinfo
  {author} {\bibfnamefont {M.~C.}\ \bibnamefont {Rechtsman}}, \bibinfo {author}
  {\bibfnamefont {Y.~D.}\ \bibnamefont {Chong}}, \bibinfo {author}
  {\bibfnamefont {M.}~\bibnamefont {Khajavikhan}}, \bibinfo {author}
  {\bibfnamefont {D.~N.}\ \bibnamefont {Christodoulides}}, \ and\ \bibinfo
  {author} {\bibfnamefont {M.}~\bibnamefont {Segev}},\ }\href
  {http://science.sciencemag.org/content/359/6381/eaar4003} {\bibfield
  {journal} {\bibinfo  {journal} {Science}\ }\textbf {\bibinfo {volume}
  {359}},\ \bibinfo {pages} {eaar4003} (\bibinfo {year} {2018})}\BibitemShut
  {NoStop}%
\bibitem [{\citenamefont {Lee}\ \emph {et~al.}(2019)\citenamefont {Lee},
  \citenamefont {Ahn}, \citenamefont {Zhou},\ and\ \citenamefont
  {Vishwanath}}]{PhysRevLett.123.206404}%
  \BibitemOpen
  \bibfield  {author} {\bibinfo {author} {\bibfnamefont {J.~Y.}\ \bibnamefont
  {Lee}}, \bibinfo {author} {\bibfnamefont {J.}~\bibnamefont {Ahn}}, \bibinfo
  {author} {\bibfnamefont {H.}~\bibnamefont {Zhou}}, \ and\ \bibinfo {author}
  {\bibfnamefont {A.}~\bibnamefont {Vishwanath}},\ }\href {\doibase
  10.1103/PhysRevLett.123.206404} {\bibfield  {journal} {\bibinfo  {journal}
  {Phys. Rev. Lett.}\ }\textbf {\bibinfo {volume} {123}},\ \bibinfo {pages}
  {206404} (\bibinfo {year} {2019})}\BibitemShut {NoStop}%
\bibitem [{\citenamefont {Borgnia}\ \emph {et~al.}(2020)\citenamefont
  {Borgnia}, \citenamefont {Kruchkov},\ and\ \citenamefont
  {Slager}}]{PhysRevLett.124.056802}%
  \BibitemOpen
  \bibfield  {author} {\bibinfo {author} {\bibfnamefont {D.~S.}\ \bibnamefont
  {Borgnia}}, \bibinfo {author} {\bibfnamefont {A.~J.}\ \bibnamefont
  {Kruchkov}}, \ and\ \bibinfo {author} {\bibfnamefont {R.-J.}\ \bibnamefont
  {Slager}},\ }\href {\doibase 10.1103/PhysRevLett.124.056802} {\bibfield
  {journal} {\bibinfo  {journal} {Phys. Rev. Lett.}\ }\textbf {\bibinfo
  {volume} {124}},\ \bibinfo {pages} {056802} (\bibinfo {year}
  {2020})}\BibitemShut {NoStop}%
\bibitem [{\citenamefont {Chen}\ and\ \citenamefont
  {Zhai}(2018)}]{PhysRevB.98.245130}%
  \BibitemOpen
  \bibfield  {author} {\bibinfo {author} {\bibfnamefont {Y.}~\bibnamefont
  {Chen}}\ and\ \bibinfo {author} {\bibfnamefont {H.}~\bibnamefont {Zhai}},\
  }\href {\doibase 10.1103/PhysRevB.98.245130} {\bibfield  {journal} {\bibinfo
  {journal} {Phys. Rev. B}\ }\textbf {\bibinfo {volume} {98}},\ \bibinfo
  {pages} {245130} (\bibinfo {year} {2018})}\BibitemShut {NoStop}%
\bibitem [{\citenamefont {Wang}\ \emph {et~al.}(2018)\citenamefont {Wang},
  \citenamefont {Zhang},\ and\ \citenamefont {Song}}]{PhysRevA.98.042120}%
  \BibitemOpen
  \bibfield  {author} {\bibinfo {author} {\bibfnamefont {R.}~\bibnamefont
  {Wang}}, \bibinfo {author} {\bibfnamefont {X.~Z.}\ \bibnamefont {Zhang}}, \
  and\ \bibinfo {author} {\bibfnamefont {Z.}~\bibnamefont {Song}},\ }\href
  {\doibase 10.1103/PhysRevA.98.042120} {\bibfield  {journal} {\bibinfo
  {journal} {Phys. Rev. A}\ }\textbf {\bibinfo {volume} {98}},\ \bibinfo
  {pages} {042120} (\bibinfo {year} {2018})}\BibitemShut {NoStop}%
\bibitem [{\citenamefont {Gong}\ \emph {et~al.}(2018)\citenamefont {Gong},
  \citenamefont {Ashida}, \citenamefont {Kawabata}, \citenamefont {Takasan},
  \citenamefont {Higashikawa},\ and\ \citenamefont {Ueda}}]{PhysRevX.8.031079}%
  \BibitemOpen
  \bibfield  {author} {\bibinfo {author} {\bibfnamefont {Z.}~\bibnamefont
  {Gong}}, \bibinfo {author} {\bibfnamefont {Y.}~\bibnamefont {Ashida}},
  \bibinfo {author} {\bibfnamefont {K.}~\bibnamefont {Kawabata}}, \bibinfo
  {author} {\bibfnamefont {K.}~\bibnamefont {Takasan}}, \bibinfo {author}
  {\bibfnamefont {S.}~\bibnamefont {Higashikawa}}, \ and\ \bibinfo {author}
  {\bibfnamefont {M.}~\bibnamefont {Ueda}},\ }\href {\doibase
  10.1103/PhysRevX.8.031079} {\bibfield  {journal} {\bibinfo  {journal} {Phys.
  Rev. X}\ }\textbf {\bibinfo {volume} {8}},\ \bibinfo {pages} {031079}
  (\bibinfo {year} {2018})}\BibitemShut {NoStop}%
\bibitem [{\citenamefont {Zhou}\ and\ \citenamefont
  {Lee}(2019)}]{PhysRevB.99.235112}%
  \BibitemOpen
  \bibfield  {author} {\bibinfo {author} {\bibfnamefont {H.}~\bibnamefont
  {Zhou}}\ and\ \bibinfo {author} {\bibfnamefont {J.~Y.}\ \bibnamefont {Lee}},\
  }\href {\doibase 10.1103/PhysRevB.99.235112} {\bibfield  {journal} {\bibinfo
  {journal} {Phys. Rev. B}\ }\textbf {\bibinfo {volume} {99}},\ \bibinfo
  {pages} {235112} (\bibinfo {year} {2019})}\BibitemShut {NoStop}%
\bibitem [{\citenamefont {Kawabata}\ \emph {et~al.}(2019)\citenamefont
  {Kawabata}, \citenamefont {Shiozaki}, \citenamefont {Ueda},\ and\
  \citenamefont {Sato}}]{PhysRevX.9.041015}%
  \BibitemOpen
  \bibfield  {author} {\bibinfo {author} {\bibfnamefont {K.}~\bibnamefont
  {Kawabata}}, \bibinfo {author} {\bibfnamefont {K.}~\bibnamefont {Shiozaki}},
  \bibinfo {author} {\bibfnamefont {M.}~\bibnamefont {Ueda}}, \ and\ \bibinfo
  {author} {\bibfnamefont {M.}~\bibnamefont {Sato}},\ }\href {\doibase
  10.1103/PhysRevX.9.041015} {\bibfield  {journal} {\bibinfo  {journal} {Phys.
  Rev. X}\ }\textbf {\bibinfo {volume} {9}},\ \bibinfo {pages} {041015}
  (\bibinfo {year} {2019})}\BibitemShut {NoStop}%
\bibitem [{\citenamefont {Yao}\ and\ \citenamefont
  {Wang}(2018)}]{PhysRevLett.121.086803}%
  \BibitemOpen
  \bibfield  {author} {\bibinfo {author} {\bibfnamefont {S.}~\bibnamefont
  {Yao}}\ and\ \bibinfo {author} {\bibfnamefont {Z.}~\bibnamefont {Wang}},\
  }\href {\doibase 10.1103/PhysRevLett.121.086803} {\bibfield  {journal}
  {\bibinfo  {journal} {Phys. Rev. Lett.}\ }\textbf {\bibinfo {volume} {121}},\
  \bibinfo {pages} {086803} (\bibinfo {year} {2018})}\BibitemShut {NoStop}%
\bibitem [{\citenamefont {Yao}\ \emph {et~al.}(2018)\citenamefont {Yao},
  \citenamefont {Song},\ and\ \citenamefont {Wang}}]{PhysRevLett.121.136802}%
  \BibitemOpen
  \bibfield  {author} {\bibinfo {author} {\bibfnamefont {S.}~\bibnamefont
  {Yao}}, \bibinfo {author} {\bibfnamefont {F.}~\bibnamefont {Song}}, \ and\
  \bibinfo {author} {\bibfnamefont {Z.}~\bibnamefont {Wang}},\ }\href
  {https://link.aps.org/doi/10.1103/PhysRevLett.121.136802} {\bibfield
  {journal} {\bibinfo  {journal} {Phys. Rev. Lett.}\ }\textbf {\bibinfo
  {volume} {121}},\ \bibinfo {pages} {136802} (\bibinfo {year}
  {2018})}\BibitemShut {NoStop}%
\bibitem [{\citenamefont {H\"ockendorf}\ \emph
  {et~al.}(2019{\natexlab{a}})\citenamefont {H\"ockendorf}, \citenamefont
  {Alvermann},\ and\ \citenamefont {Fehske}}]{PhysRevLett.123.190403}%
  \BibitemOpen
  \bibfield  {author} {\bibinfo {author} {\bibfnamefont {B.}~\bibnamefont
  {H\"ockendorf}}, \bibinfo {author} {\bibfnamefont {A.}~\bibnamefont
  {Alvermann}}, \ and\ \bibinfo {author} {\bibfnamefont {H.}~\bibnamefont
  {Fehske}},\ }\href {\doibase 10.1103/PhysRevLett.123.190403} {\bibfield
  {journal} {\bibinfo  {journal} {Phys. Rev. Lett.}\ }\textbf {\bibinfo
  {volume} {123}},\ \bibinfo {pages} {190403} (\bibinfo {year}
  {2019}{\natexlab{a}})}\BibitemShut {NoStop}%
\bibitem [{\citenamefont {Lee}\ and\ \citenamefont
  {Thomale}(2019)}]{PhysRevB.99.201103}%
  \BibitemOpen
  \bibfield  {author} {\bibinfo {author} {\bibfnamefont {C.~H.}\ \bibnamefont
  {Lee}}\ and\ \bibinfo {author} {\bibfnamefont {R.}~\bibnamefont {Thomale}},\
  }\href {\doibase 10.1103/PhysRevB.99.201103} {\bibfield  {journal} {\bibinfo
  {journal} {Phys. Rev. B}\ }\textbf {\bibinfo {volume} {99}},\ \bibinfo
  {pages} {201103} (\bibinfo {year} {2019})}\BibitemShut {NoStop}%
\bibitem [{\citenamefont {Zhu}\ \emph {et~al.}(2018)\citenamefont {Zhu},
  \citenamefont {Fang}, \citenamefont {Li}, \citenamefont {Sun}, \citenamefont
  {Li}, \citenamefont {Jing},\ and\ \citenamefont
  {Chen}}]{PhysRevLett.121.124501}%
  \BibitemOpen
  \bibfield  {author} {\bibinfo {author} {\bibfnamefont {W.}~\bibnamefont
  {Zhu}}, \bibinfo {author} {\bibfnamefont {X.}~\bibnamefont {Fang}}, \bibinfo
  {author} {\bibfnamefont {D.}~\bibnamefont {Li}}, \bibinfo {author}
  {\bibfnamefont {Y.}~\bibnamefont {Sun}}, \bibinfo {author} {\bibfnamefont
  {Y.}~\bibnamefont {Li}}, \bibinfo {author} {\bibfnamefont {Y.}~\bibnamefont
  {Jing}}, \ and\ \bibinfo {author} {\bibfnamefont {H.}~\bibnamefont {Chen}},\
  }\href {\doibase 10.1103/PhysRevLett.121.124501} {\bibfield  {journal}
  {\bibinfo  {journal} {Phys. Rev. Lett.}\ }\textbf {\bibinfo {volume} {121}},\
  \bibinfo {pages} {124501} (\bibinfo {year} {2018})}\BibitemShut {NoStop}%
\bibitem [{\citenamefont {{Helbig}}\ \emph {et~al.}(2019)\citenamefont
  {{Helbig}}, \citenamefont {{Hofmann}}, \citenamefont {{Imhof}}, \citenamefont
  {{Abdelghany}}, \citenamefont {{Kiessling}}, \citenamefont {{Molenkamp}},
  \citenamefont {{Lee}}, \citenamefont {{Szameit}}, \citenamefont {{Greiter}},\
  and\ \citenamefont {{Thomale}}}]{2019arXiv190711562H}%
  \BibitemOpen
  \bibfield  {author} {\bibinfo {author} {\bibfnamefont {T.}~\bibnamefont
  {{Helbig}}}, \bibinfo {author} {\bibfnamefont {T.}~\bibnamefont {{Hofmann}}},
  \bibinfo {author} {\bibfnamefont {S.}~\bibnamefont {{Imhof}}}, \bibinfo
  {author} {\bibfnamefont {M.}~\bibnamefont {{Abdelghany}}}, \bibinfo {author}
  {\bibfnamefont {T.}~\bibnamefont {{Kiessling}}}, \bibinfo {author}
  {\bibfnamefont {L.~W.}\ \bibnamefont {{Molenkamp}}}, \bibinfo {author}
  {\bibfnamefont {C.~H.}\ \bibnamefont {{Lee}}}, \bibinfo {author}
  {\bibfnamefont {A.}~\bibnamefont {{Szameit}}}, \bibinfo {author}
  {\bibfnamefont {M.}~\bibnamefont {{Greiter}}}, \ and\ \bibinfo {author}
  {\bibfnamefont {R.}~\bibnamefont {{Thomale}}},\ }\href
  {https://arxiv.org/abs/1907.11562} {\bibfield  {journal} {\bibinfo  {journal}
  {arXiv:1907.11562}\ } (\bibinfo {year} {2019})}\BibitemShut {NoStop}%
\bibitem [{\citenamefont {Szameit}\ and\ \citenamefont
  {Nolte}(2010)}]{SzameitJPB}%
  \BibitemOpen
  \bibfield  {author} {\bibinfo {author} {\bibfnamefont {A.}~\bibnamefont
  {Szameit}}\ and\ \bibinfo {author} {\bibfnamefont {S.}~\bibnamefont
  {Nolte}},\ }\href {http://stacks.iop.org/0953-4075/43/i=16/a=163001}
  {\bibfield  {journal} {\bibinfo  {journal} {J. Phys. B}\ }\textbf {\bibinfo
  {volume} {43}},\ \bibinfo {pages} {163001} (\bibinfo {year}
  {2010})}\BibitemShut {NoStop}%
\bibitem [{\citenamefont {Ozawa}\ \emph {et~al.}(2019)\citenamefont {Ozawa},
  \citenamefont {Price}, \citenamefont {Amo}, \citenamefont {Goldman},
  \citenamefont {Hafezi}, \citenamefont {Lu}, \citenamefont {Rechtsman},
  \citenamefont {Schuster}, \citenamefont {Simon}, \citenamefont {Zilberberg},\
  and\ \citenamefont {Carusotto}}]{RevModPhys.91.015006}%
  \BibitemOpen
  \bibfield  {author} {\bibinfo {author} {\bibfnamefont {T.}~\bibnamefont
  {Ozawa}}, \bibinfo {author} {\bibfnamefont {H.~M.}\ \bibnamefont {Price}},
  \bibinfo {author} {\bibfnamefont {A.}~\bibnamefont {Amo}}, \bibinfo {author}
  {\bibfnamefont {N.}~\bibnamefont {Goldman}}, \bibinfo {author} {\bibfnamefont
  {M.}~\bibnamefont {Hafezi}}, \bibinfo {author} {\bibfnamefont
  {L.}~\bibnamefont {Lu}}, \bibinfo {author} {\bibfnamefont {M.~C.}\
  \bibnamefont {Rechtsman}}, \bibinfo {author} {\bibfnamefont {D.}~\bibnamefont
  {Schuster}}, \bibinfo {author} {\bibfnamefont {J.}~\bibnamefont {Simon}},
  \bibinfo {author} {\bibfnamefont {O.}~\bibnamefont {Zilberberg}}, \ and\
  \bibinfo {author} {\bibfnamefont {I.}~\bibnamefont {Carusotto}},\ }\href
  {\doibase 10.1103/RevModPhys.91.015006} {\bibfield  {journal} {\bibinfo
  {journal} {Rev. Mod. Phys.}\ }\textbf {\bibinfo {volume} {91}},\ \bibinfo
  {pages} {015006} (\bibinfo {year} {2019})}\BibitemShut {NoStop}%
\bibitem [{\citenamefont {Regensburger}\ \emph {et~al.}(2012)\citenamefont
  {Regensburger}, \citenamefont {Bersch}, \citenamefont {Miri}, \citenamefont
  {Onishchukov}, \citenamefont {Christodoulides},\ and\ \citenamefont
  {Peschel}}]{Regensburger2012}%
  \BibitemOpen
  \bibfield  {author} {\bibinfo {author} {\bibfnamefont {A.}~\bibnamefont
  {Regensburger}}, \bibinfo {author} {\bibfnamefont {C.}~\bibnamefont
  {Bersch}}, \bibinfo {author} {\bibfnamefont {M.-A.}\ \bibnamefont {Miri}},
  \bibinfo {author} {\bibfnamefont {G.}~\bibnamefont {Onishchukov}}, \bibinfo
  {author} {\bibfnamefont {D.~N.}\ \bibnamefont {Christodoulides}}, \ and\
  \bibinfo {author} {\bibfnamefont {U.}~\bibnamefont {Peschel}},\ }\href
  {https://doi.org/10.1038/nature11298} {\bibfield  {journal} {\bibinfo
  {journal} {Nature}\ }\textbf {\bibinfo {volume} {488}},\ \bibinfo {pages}
  {167} (\bibinfo {year} {2012})}\BibitemShut {NoStop}%
\bibitem [{\citenamefont {Weimann}\ \emph {et~al.}(2016)\citenamefont
  {Weimann}, \citenamefont {Kremer}, \citenamefont {Plotnik}, \citenamefont
  {Lumer}, \citenamefont {Nolte}, \citenamefont {Makris}, \citenamefont
  {Segev}, \citenamefont {Rechtsman},\ and\ \citenamefont
  {Szameit}}]{Weimann2016}%
  \BibitemOpen
  \bibfield  {author} {\bibinfo {author} {\bibfnamefont {S.}~\bibnamefont
  {Weimann}}, \bibinfo {author} {\bibfnamefont {M.}~\bibnamefont {Kremer}},
  \bibinfo {author} {\bibfnamefont {Y.}~\bibnamefont {Plotnik}}, \bibinfo
  {author} {\bibfnamefont {Y.}~\bibnamefont {Lumer}}, \bibinfo {author}
  {\bibfnamefont {S.}~\bibnamefont {Nolte}}, \bibinfo {author} {\bibfnamefont
  {K.~G.}\ \bibnamefont {Makris}}, \bibinfo {author} {\bibfnamefont
  {M.}~\bibnamefont {Segev}}, \bibinfo {author} {\bibfnamefont {M.~C.}\
  \bibnamefont {Rechtsman}}, \ and\ \bibinfo {author} {\bibfnamefont
  {A.}~\bibnamefont {Szameit}},\ }\href {https://doi.org/10.1038/nmat4811}
  {\bibfield  {journal} {\bibinfo  {journal} {Nat. Mat.}\ }\textbf {\bibinfo
  {volume} {16}},\ \bibinfo {pages} {433} (\bibinfo {year} {2016})}\BibitemShut
  {NoStop}%
\bibitem [{\citenamefont {Zeuner}\ \emph {et~al.}(2015)\citenamefont {Zeuner},
  \citenamefont {Rechtsman}, \citenamefont {Plotnik}, \citenamefont {Lumer},
  \citenamefont {Nolte}, \citenamefont {Rudner}, \citenamefont {Segev},\ and\
  \citenamefont {Szameit}}]{PhysRevLett.115.040402}%
  \BibitemOpen
  \bibfield  {author} {\bibinfo {author} {\bibfnamefont {J.~M.}\ \bibnamefont
  {Zeuner}}, \bibinfo {author} {\bibfnamefont {M.~C.}\ \bibnamefont
  {Rechtsman}}, \bibinfo {author} {\bibfnamefont {Y.}~\bibnamefont {Plotnik}},
  \bibinfo {author} {\bibfnamefont {Y.}~\bibnamefont {Lumer}}, \bibinfo
  {author} {\bibfnamefont {S.}~\bibnamefont {Nolte}}, \bibinfo {author}
  {\bibfnamefont {M.~S.}\ \bibnamefont {Rudner}}, \bibinfo {author}
  {\bibfnamefont {M.}~\bibnamefont {Segev}}, \ and\ \bibinfo {author}
  {\bibfnamefont {A.}~\bibnamefont {Szameit}},\ }\href
  {https://link.aps.org/doi/10.1103/PhysRevLett.115.040402} {\bibfield
  {journal} {\bibinfo  {journal} {Phys. Rev. Lett.}\ }\textbf {\bibinfo
  {volume} {115}},\ \bibinfo {pages} {040402} (\bibinfo {year}
  {2015})}\BibitemShut {NoStop}%
\bibitem [{\citenamefont {Bandres}\ \emph {et~al.}(2018)\citenamefont
  {Bandres}, \citenamefont {Wittek}, \citenamefont {Harari}, \citenamefont
  {Parto}, \citenamefont {Ren}, \citenamefont {Segev}, \citenamefont
  {Christodoulides},\ and\ \citenamefont {Khajavikhan}}]{bandres2018}%
  \BibitemOpen
  \bibfield  {author} {\bibinfo {author} {\bibfnamefont {M.~A.}\ \bibnamefont
  {Bandres}}, \bibinfo {author} {\bibfnamefont {S.}~\bibnamefont {Wittek}},
  \bibinfo {author} {\bibfnamefont {G.}~\bibnamefont {Harari}}, \bibinfo
  {author} {\bibfnamefont {M.}~\bibnamefont {Parto}}, \bibinfo {author}
  {\bibfnamefont {J.}~\bibnamefont {Ren}}, \bibinfo {author} {\bibfnamefont
  {M.}~\bibnamefont {Segev}}, \bibinfo {author} {\bibfnamefont {D.~N.}\
  \bibnamefont {Christodoulides}}, \ and\ \bibinfo {author} {\bibfnamefont
  {M.}~\bibnamefont {Khajavikhan}},\ }\href {\doibase 10.1126/science.aar4005}
  {\bibfield  {journal} {\bibinfo  {journal} {Science}\ }\textbf {\bibinfo
  {volume} {359}},\ \bibinfo {pages} {eaar4005} (\bibinfo {year}
  {2018})}\BibitemShut {NoStop}%
\bibitem [{\citenamefont {Kitagawa}\ \emph {et~al.}(2010)\citenamefont
  {Kitagawa}, \citenamefont {Berg}, \citenamefont {Rudner},\ and\ \citenamefont
  {Demler}}]{KitagawaPRB}%
  \BibitemOpen
  \bibfield  {author} {\bibinfo {author} {\bibfnamefont {T.}~\bibnamefont
  {Kitagawa}}, \bibinfo {author} {\bibfnamefont {E.}~\bibnamefont {Berg}},
  \bibinfo {author} {\bibfnamefont {M.}~\bibnamefont {Rudner}}, \ and\ \bibinfo
  {author} {\bibfnamefont {E.}~\bibnamefont {Demler}},\ }\href
  {http://link.aps.org/doi/10.1103/PhysRevB.82.235114} {\bibfield  {journal}
  {\bibinfo  {journal} {Phys. Rev. B}\ }\textbf {\bibinfo {volume} {82}},\
  \bibinfo {pages} {235114} (\bibinfo {year} {2010})}\BibitemShut {NoStop}%
\bibitem [{\citenamefont {Rudner}\ \emph {et~al.}(2013)\citenamefont {Rudner},
  \citenamefont {Lindner}, \citenamefont {Berg},\ and\ \citenamefont
  {Levin}}]{Rudner}%
  \BibitemOpen
  \bibfield  {author} {\bibinfo {author} {\bibfnamefont {M.~S.}\ \bibnamefont
  {Rudner}}, \bibinfo {author} {\bibfnamefont {N.~H.}\ \bibnamefont {Lindner}},
  \bibinfo {author} {\bibfnamefont {E.}~\bibnamefont {Berg}}, \ and\ \bibinfo
  {author} {\bibfnamefont {M.}~\bibnamefont {Levin}},\ }\href
  {http://link.aps.org/doi/10.1103/PhysRevX.3.031005} {\bibfield  {journal}
  {\bibinfo  {journal} {Phys. Rev. X}\ }\textbf {\bibinfo {volume} {3}},\
  \bibinfo {pages} {031005} (\bibinfo {year} {2013})}\BibitemShut {NoStop}%
\bibitem [{\citenamefont {Nathan}\ and\ \citenamefont {Rudner}(2015)}]{Nathan}%
  \BibitemOpen
  \bibfield  {author} {\bibinfo {author} {\bibfnamefont {F.}~\bibnamefont
  {Nathan}}\ and\ \bibinfo {author} {\bibfnamefont {M.~S.}\ \bibnamefont
  {Rudner}},\ }\href {http://stacks.iop.org/1367-2630/17/i=12/a=125014}
  {\bibfield  {journal} {\bibinfo  {journal} {New J. Phys.}\ }\textbf {\bibinfo
  {volume} {17}},\ \bibinfo {pages} {125014} (\bibinfo {year}
  {2015})}\BibitemShut {NoStop}%
\bibitem [{\citenamefont {H{\"o}ckendorf}\ \emph {et~al.}(2017)\citenamefont
  {H{\"o}ckendorf}, \citenamefont {Alvermann},\ and\ \citenamefont
  {Fehske}}]{HockendorfJPA}%
  \BibitemOpen
  \bibfield  {author} {\bibinfo {author} {\bibfnamefont {B.}~\bibnamefont
  {H{\"o}ckendorf}}, \bibinfo {author} {\bibfnamefont {A.}~\bibnamefont
  {Alvermann}}, \ and\ \bibinfo {author} {\bibfnamefont {H.}~\bibnamefont
  {Fehske}},\ }\href {http://stacks.iop.org/1751-8121/50/i=29/a=295301}
  {\bibfield  {journal} {\bibinfo  {journal} {J. Phys. A}\ }\textbf {\bibinfo
  {volume} {50}},\ \bibinfo {pages} {295301} (\bibinfo {year}
  {2017})}\BibitemShut {NoStop}%
\bibitem [{\citenamefont {H\"ockendorf}\ \emph {et~al.}(2018)\citenamefont
  {H\"ockendorf}, \citenamefont {Alvermann},\ and\ \citenamefont
  {Fehske}}]{HockendorfPRB}%
  \BibitemOpen
  \bibfield  {author} {\bibinfo {author} {\bibfnamefont {B.}~\bibnamefont
  {H\"ockendorf}}, \bibinfo {author} {\bibfnamefont {A.}~\bibnamefont
  {Alvermann}}, \ and\ \bibinfo {author} {\bibfnamefont {H.}~\bibnamefont
  {Fehske}},\ }\href {\doibase 10.1103/PhysRevB.97.045140} {\bibfield
  {journal} {\bibinfo  {journal} {Phys. Rev. B}\ }\textbf {\bibinfo {volume}
  {97}},\ \bibinfo {pages} {045140} (\bibinfo {year} {2018})}\BibitemShut
  {NoStop}%
\bibitem [{\citenamefont {H\"ockendorf}\ \emph
  {et~al.}(2019{\natexlab{b}})\citenamefont {H\"ockendorf}, \citenamefont
  {Alvermann},\ and\ \citenamefont {Fehske}}]{HAF19}%
  \BibitemOpen
  \bibfield  {author} {\bibinfo {author} {\bibfnamefont {B.}~\bibnamefont
  {H\"ockendorf}}, \bibinfo {author} {\bibfnamefont {A.}~\bibnamefont
  {Alvermann}}, \ and\ \bibinfo {author} {\bibfnamefont {H.}~\bibnamefont
  {Fehske}},\ }\href {\doibase 10.1103/PhysRevB.99.245102} {\bibfield
  {journal} {\bibinfo  {journal} {Phys. Rev. B}\ }\textbf {\bibinfo {volume}
  {99}},\ \bibinfo {pages} {245102} (\bibinfo {year}
  {2019}{\natexlab{b}})}\BibitemShut {NoStop}%
\bibitem [{\citenamefont {Maczewsky}\ \emph {et~al.}(2017)\citenamefont
  {Maczewsky}, \citenamefont {Zeuner}, \citenamefont {Nolte},\ and\
  \citenamefont {Szameit}}]{Maczewsky}%
  \BibitemOpen
  \bibfield  {author} {\bibinfo {author} {\bibfnamefont {L.~J.}\ \bibnamefont
  {Maczewsky}}, \bibinfo {author} {\bibfnamefont {J.~M.}\ \bibnamefont
  {Zeuner}}, \bibinfo {author} {\bibfnamefont {S.}~\bibnamefont {Nolte}}, \
  and\ \bibinfo {author} {\bibfnamefont {A.}~\bibnamefont {Szameit}},\ }\href
  {http://dx.doi.org/10.1038/ncomms13756} {\bibfield  {journal} {\bibinfo
  {journal} {Nat. Comm.}\ }\textbf {\bibinfo {volume} {8}},\ \bibinfo {pages}
  {13756} (\bibinfo {year} {2017})}\BibitemShut {NoStop}%
\bibitem [{\citenamefont {Mukherjee}\ \emph {et~al.}(2017)\citenamefont
  {Mukherjee}, \citenamefont {Spracklen}, \citenamefont {Valiente},
  \citenamefont {Andersson}, \citenamefont {{\"O}hberg}, \citenamefont
  {Goldman},\ and\ \citenamefont {Thomson}}]{Mukherjee}%
  \BibitemOpen
  \bibfield  {author} {\bibinfo {author} {\bibfnamefont {S.}~\bibnamefont
  {Mukherjee}}, \bibinfo {author} {\bibfnamefont {A.}~\bibnamefont
  {Spracklen}}, \bibinfo {author} {\bibfnamefont {M.}~\bibnamefont {Valiente}},
  \bibinfo {author} {\bibfnamefont {E.}~\bibnamefont {Andersson}}, \bibinfo
  {author} {\bibfnamefont {P.}~\bibnamefont {{\"O}hberg}}, \bibinfo {author}
  {\bibfnamefont {N.}~\bibnamefont {Goldman}}, \ and\ \bibinfo {author}
  {\bibfnamefont {R.~R.}\ \bibnamefont {Thomson}},\ }\href
  {http://dx.doi.org/10.1038/ncomms13918} {\bibfield  {journal} {\bibinfo
  {journal} {Nat. Comm.}\ }\textbf {\bibinfo {volume} {8}},\ \bibinfo {pages}
  {13918} (\bibinfo {year} {2017})}\BibitemShut {NoStop}%
\bibitem [{\citenamefont {Peng}\ \emph {et~al.}(2016)\citenamefont {Peng},
  \citenamefont {Qin}, \citenamefont {Zhao}, \citenamefont {Shen},
  \citenamefont {Xu}, \citenamefont {Bao}, \citenamefont {Jia},\ and\
  \citenamefont {Zhu}}]{Peng2016}%
  \BibitemOpen
  \bibfield  {author} {\bibinfo {author} {\bibfnamefont {Y.-G.}\ \bibnamefont
  {Peng}}, \bibinfo {author} {\bibfnamefont {C.-Z.}\ \bibnamefont {Qin}},
  \bibinfo {author} {\bibfnamefont {D.-G.}\ \bibnamefont {Zhao}}, \bibinfo
  {author} {\bibfnamefont {Y.-X.}\ \bibnamefont {Shen}}, \bibinfo {author}
  {\bibfnamefont {X.-Y.}\ \bibnamefont {Xu}}, \bibinfo {author} {\bibfnamefont
  {M.}~\bibnamefont {Bao}}, \bibinfo {author} {\bibfnamefont {H.}~\bibnamefont
  {Jia}}, \ and\ \bibinfo {author} {\bibfnamefont {X.-F.}\ \bibnamefont
  {Zhu}},\ }\href {https://doi.org/10.1038/ncomms13368} {\bibfield  {journal}
  {\bibinfo  {journal} {Nat. Comm.}\ }\textbf {\bibinfo {volume} {7}},\
  \bibinfo {pages} {13368} (\bibinfo {year} {2016})}\BibitemShut {NoStop}%
\bibitem [{\citenamefont {{Maczewsky}}\ \emph {et~al.}(2018)\citenamefont
  {{Maczewsky}}, \citenamefont {{H{\"o}ckendorf}}, \citenamefont {{Kremer}},
  \citenamefont {{Biesenthal}}, \citenamefont {{Heinrich}}, \citenamefont
  {{Alvermann}}, \citenamefont {{Fehske}},\ and\ \citenamefont
  {{Szameit}}}]{GreifRostock}%
  \BibitemOpen
  \bibfield  {author} {\bibinfo {author} {\bibfnamefont {L.~J.}\ \bibnamefont
  {{Maczewsky}}}, \bibinfo {author} {\bibfnamefont {B.}~\bibnamefont
  {{H{\"o}ckendorf}}}, \bibinfo {author} {\bibfnamefont {M.}~\bibnamefont
  {{Kremer}}}, \bibinfo {author} {\bibfnamefont {T.}~\bibnamefont
  {{Biesenthal}}}, \bibinfo {author} {\bibfnamefont {M.}~\bibnamefont
  {{Heinrich}}}, \bibinfo {author} {\bibfnamefont {A.}~\bibnamefont
  {{Alvermann}}}, \bibinfo {author} {\bibfnamefont {H.}~\bibnamefont
  {{Fehske}}}, \ and\ \bibinfo {author} {\bibfnamefont {A.}~\bibnamefont
  {{Szameit}}},\ }\href {https://arxiv.org/abs/1812.07930} {\bibfield
  {journal} {\bibinfo  {journal} {arXiv:1812.07930}\ } (\bibinfo {year}
  {2018})}\BibitemShut {NoStop}%
\bibitem [{\citenamefont {Graf}\ and\ \citenamefont {Tauber}(2018)}]{Graf2018}%
  \BibitemOpen
  \bibfield  {author} {\bibinfo {author} {\bibfnamefont {G.~M.}\ \bibnamefont
  {Graf}}\ and\ \bibinfo {author} {\bibfnamefont {C.}~\bibnamefont {Tauber}},\
  }\href {\doibase 10.1007/s00023-018-0657-7} {\bibfield  {journal} {\bibinfo
  {journal} {Ann. Henri Poincar{\'e}}\ }\textbf {\bibinfo {volume} {19}},\
  \bibinfo {pages} {709} (\bibinfo {year} {2018})}\BibitemShut {NoStop}%
\bibitem [{\citenamefont {Tauber}(2018)}]{PhysRevB.97.195312}%
  \BibitemOpen
  \bibfield  {author} {\bibinfo {author} {\bibfnamefont {C.}~\bibnamefont
  {Tauber}},\ }\href {\doibase 10.1103/PhysRevB.97.195312} {\bibfield
  {journal} {\bibinfo  {journal} {Phys. Rev. B}\ }\textbf {\bibinfo {volume}
  {97}},\ \bibinfo {pages} {195312} (\bibinfo {year} {2018})}\BibitemShut
  {NoStop}%
\bibitem [{\citenamefont {Golub}\ and\ \citenamefont
  {Van~Loan}(2013)}]{GolubVanLoan}%
  \BibitemOpen
  \bibfield  {author} {\bibinfo {author} {\bibfnamefont {G.~H.}\ \bibnamefont
  {Golub}}\ and\ \bibinfo {author} {\bibfnamefont {C.~F.}\ \bibnamefont
  {Van~Loan}},\ }\href@noop {} {\emph {\bibinfo {title} {Matrix
  Computations}}},\ \bibinfo {edition} {4th}\ ed.\ (\bibinfo  {publisher} {The
  Johns Hopkins University Press},\ \bibinfo {year} {2013})\BibitemShut
  {NoStop}%
\bibitem [{\citenamefont {Fedorova}\ \emph {et~al.}(2019)\citenamefont
  {Fedorova}, \citenamefont {Qiu}, \citenamefont {Linden},\ and\ \citenamefont
  {Kroha}}]{fedorova2019topological}%
  \BibitemOpen
  \bibfield  {author} {\bibinfo {author} {\bibfnamefont {Z.}~\bibnamefont
  {Fedorova}}, \bibinfo {author} {\bibfnamefont {H.}~\bibnamefont {Qiu}},
  \bibinfo {author} {\bibfnamefont {S.}~\bibnamefont {Linden}}, \ and\ \bibinfo
  {author} {\bibfnamefont {J.}~\bibnamefont {Kroha}},\ }\href
  {https://arxiv.org/abs/1911.03770} {\bibfield  {journal} {\bibinfo  {journal}
  {arXiv:1911.03770}\ } (\bibinfo {year} {2019})}\BibitemShut {NoStop}%
\bibitem [{\citenamefont {Abramowitz}\ and\ \citenamefont
  {Stegun}(1970)}]{AS70}%
  \BibitemOpen
  \bibinfo {editor} {\bibfnamefont {M.}~\bibnamefont {Abramowitz}}\ and\
  \bibinfo {editor} {\bibfnamefont {I.~A.}\ \bibnamefont {Stegun}},\ eds.,\
  \href@noop {} {\emph {\bibinfo {title} {{Handbook} of {Mathematical}
  {Functions} with {Formulas}, {Graphs}, and {Mathematical} {Tables}}}}\
  (\bibinfo  {publisher} {Dover Publications},\ \bibinfo {year}
  {1970})\BibitemShut {NoStop}%
\end{thebibliography}
\end{document}